# Model of Outgrowths in the Spiral Galaxies NGC 4921 and NGC 7049 and the Origin of Spiral Arms

Per Carlqvist

*Space and Plasma Physics, School of Electrical Engineering, Royal Institute of Technology, 100 44 Stockholm, Sweden*

**Abstract** NGC 4921 and 7049 are two spiral galaxies presenting narrow, distinct dust features. A detailed study of the morphology of those features has been carried out using Hubble Space Telescope archival images[1]. NGC 4921 shows a few but well-defined dust arms midway to its centre while NGC 7049 displays many more dusty features, mainly collected within a ring-shaped formation. Numerous dark and filamentary structures, called *outgrowths,* are found to protrude from the dusty arms in both galaxies. The outgrowths point both outwards and inwards in the galaxies. Mostly they are found to be V-shaped or Y-shaped with the branches connected to dark arm filaments. Often the stem of the Y appears to consist of intertwined filaments. Remarkably, the outgrowths show considerable similarities to elephant trunks in H II regions. A model of the outgrowths, based on magnetized filaments, is proposed. The model provides explanations of both the shapes and orientations of the outgrowths. Most important, it can also give an account for their intertwined structures. It is found that the longest outgrowths are confusingly similar to dusty spiral arms. This suggests that some of the outgrowths can develop into such arms. The time-scale of the development is estimated to be on the order of the rotation period of the arms or shorter. Similar processes may also take place in other spiral galaxies. If so, the model of the outgrowths can offer a new approach to the old winding problem of spiral arms.

## 1 Introduction

Galaxies containing dusty features can sometimes display remarkable and unexpected shapes. An illustration of that is the giant, elliptical galaxy NGC 1316 in Fornax showing a conspicuous web of dusty filaments (Evans 1949; Shklovskii and Cholopov 1952; de Vaucouleurs 1953; Schweizer 1980; Grillmair et al. 1999). It is believed that the galaxy is a merger, which has captured one or more gas-rich galaxies during the last three billion years (Schweizer 1980; Horellou et al. 2001; Goudfrooij et al. 2004). An intriguing feature of NGC

[1] Based on observations made with the NASA/ESA Hubble Space Telescope, obtained from the Data Archive at the Space Telescope Science Institute, which is operated by the Association of Universities for Research in Astronomy, Inc., under NASA contract NAS 5-26555. These observations are associated with programs ID 10842 and ID 9427.

1316 is constituted by a great number of dark structures protruding from the dusty filaments (Carlqvist 2010). The structures are all filamentary and often twisted. Remarkably enough, the protruding structures are oriented only inwards in the galaxy, although with a considerable scattering. The fact that not a single one of these structures is apparently oriented outwards makes them particularly challenging.

A closer study of NGC 1316 shows that the protruding structures are in several respects very similar to elephant trunks in H II regions. Only their sizes appear to clearly differ. The structures in NGC 1316 are typically a factor of thousand larger than normal, well-developed elephant trunks. As a consequence of that, the structures have been named, *mammoth trunks*. A model of the mammoth trunks, based on magnetized filaments, has recently been proposed (Carlqvist 2010). The model cannot only account for the filamentary and twisted structures of the trunks but also offers an explanation of their inward orientations.

The fact that there exist protruding mammoth trunks in the odd-looking galaxy NGC 1316 naturally leads to the question whether there may also exist similar protruding structures in other, more regular, galaxies. In order to find out whether that may be the case, we have selected two spiral galaxies with distinct dust features, NGC 4921 and 7049, for a closer study. Although both of these galaxies show dark spiral patterns, they differ considerably from one another. NGC 4921 is a so-called anaemic galaxy with diffuse bright arms and only a few narrow, but distinct, dust arms while NGC 7049 presents a more densely packed system of dusty arms and other dark features.

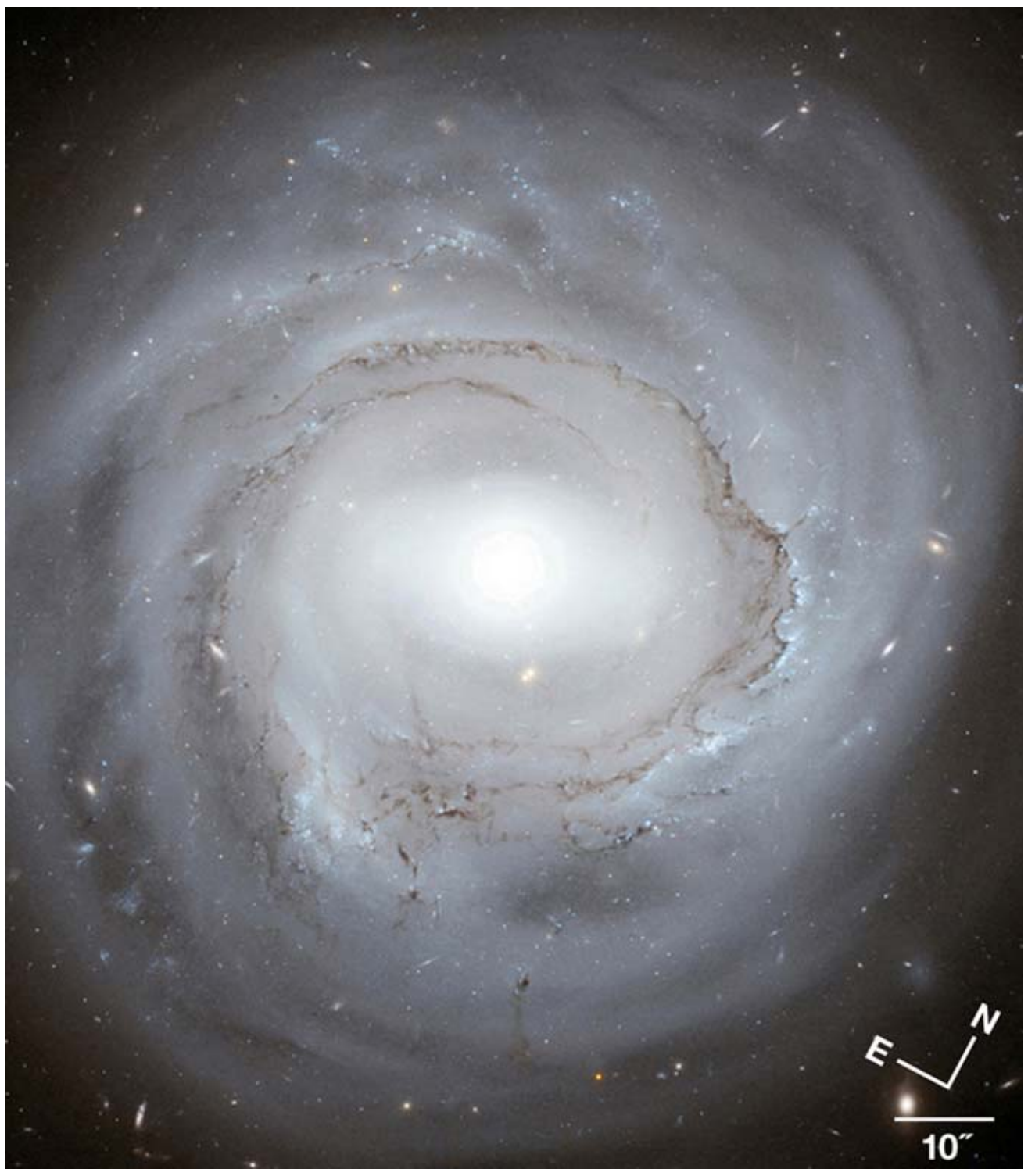

**Figure 1.** A very deep HST, *ACS* image of the spiral galaxy NGC 4921 assembled by F606W and F814W filter images (Credit: NASA, ESA, K. Cook (Lawrence Livermore National Laboratory, USA)). The outer parts of the galaxy present pale, smooth, spiral arms, which are fairly tightly wound. Halfway to the centre, a few distinct dust arms can be seen. A striking feature of the dust arms is that they form the basis of a great number of dark outgrowths, which are putting out like thorns from the arm filaments. Clusters of young, blue stars are associated with many of the outgrowths. The centre of the galaxy is at $13^h\ 1^m.4$, $+ 27° 53′$, (Epoch 2000.0). Also the more detailed images of NGC 4921 shown in Figures 2 to 7 and Figure 16 originate from the above image and are oriented in the same way.

## 2 Morphology of NGC 4921

NGC 4921 is a SB(rs)ab galaxy, one of few spiral galaxies in the Coma Galaxy Cluster (Abell 1656). A recent, very deep image captured by the Hubble Space Telescope (HST), *Advanced Camera for Surveys* (*ACS*) shows NGC 4921 as a nearly round galaxy seen face-on (Figure 1). The diameter is about 2′ corresponding to ~ 60 kpc at an adopted distance of ~100 Mpc to the Coma Cluster (1″ ≈ 485 pc). The outer parts of the galaxy consist of smooth, pale, and fairly tightly wound arms. The innermost region contains a bright nucleus surrounded by a diffuse and relatively modest bar. No dusty features are to be seen in this region in Figure 1 but a high-pass filter version of the image reveals several small, dark, spiral arms inside a radius of ~ 5″.

The intermediate region of the galaxy presents quite a different view. Here, a few narrow dust arms spiral around the central region. The dust arms (also called primary dust lanes (Lynds 1970)) are very distinct and thus in striking contrasts to the rest of the galaxy. A notable feature of the dust arms is, as pointed out by Tikhonov & Galazutdinova (2011), that they are furnished with a great number of dark outgrowths. In a sense, they very much resemble the mammoth trunks in NGC 1316. At or just outside the tips of many of the outgrowths, young, blue stars and clusters of such stars are to be seen. A number of distant galaxies, situated behind NGC 4921, are clearly visible through the disc of the galaxy revealing that the pale arms are fairly transparent. Compared with dust arms in many other spiral galaxies, the distinct dust arms in NGC 4921 are very simple. This makes them and other dark features in NGC 4921 ideal objects for a closer investigation.

The dust arms mostly form tight spirals at radial distances in the interval ~15″– 40″ (~7 – 20 kpc). A remarkable property of the arms is that they are generally made up of two or more dark filaments. In some places the arms are as narrow as ~ 0.″5 – 2″. Good examples of that are the two narrow arms shown in the lower half of Figure 2. The arms consist of at least two dark filaments, which in long pieces appear to be twined around each other. Typical widths of the filaments are found in the interval 0.″10 – 0.″25. Also on other locations in NGC 4921 the dust arms are divided into two or more filaments being locally intertwined. In many places dark outgrowths are jutting out from the arm filaments. In contrast to the situation in NGC 1316, the outgrowths are found to be oriented *both outwards and inwards* in NGC 4921. Clusters of hot, young stars are spread outside most of the outgrowths.

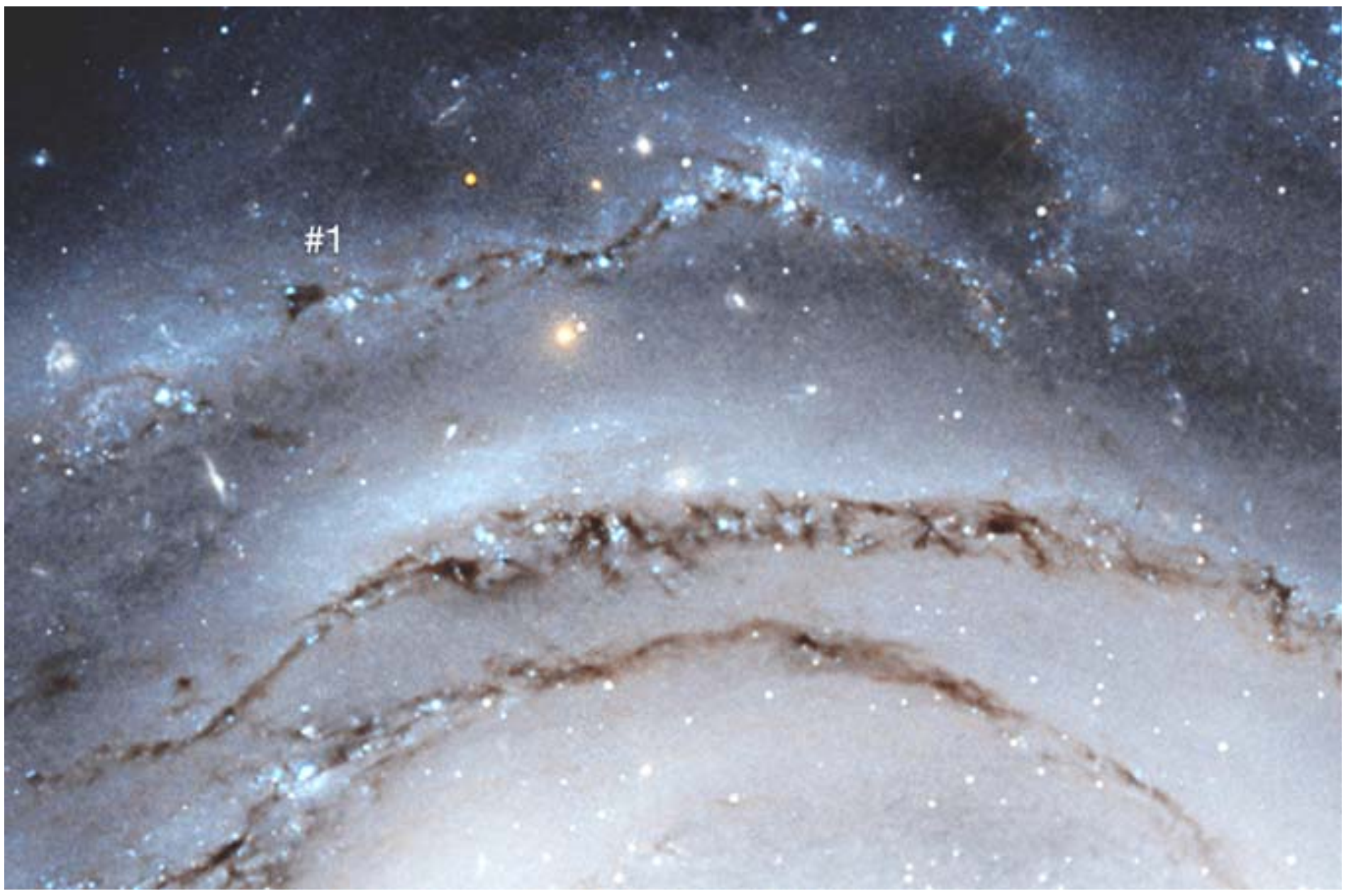

**Figure 2.** Dark features in the north-eastern quadrant of NGC 4921. In the lower half of the image two distinct dust arms are present. Each of them consists of at least two narrow filaments, which in places appear to be wound around one another. In the upper half of the image, an arm-like, dark structure (#1) sprinkled with young, blue stars is to be seen. The structure is attached to the outer dust arm by means of two or more comparatively faint legs situated near the right-hand edge of the image. A massive head finishes the structure to the left. Image size: 42″ x 28″.

About 25″ north-north-west of the centre of the galaxy there are three outgrowths of lengths ~1″– 2″ (~ 0.5 – 1 kpc) oriented outwards from the outer dust arm (Figure 3). All of them are filamentary and have diverging legs connecting them with arm filaments at their bases. The outgrowths form straight pillars with sub-structures suggesting that some of the filaments are twisted about each other. Young, blue stars are present near the tips, or heads, of the outgrowths. A few arc seconds to the west of the three outgrowths is a bulge, which at first sight shows little similarity to the three outgrowths. The presence of a few blue stars outside the bulge, however, tells us that it may once have been a larger outgrowth. This view is also supported by the fact that there is a thin and wavy filament protruding from the top of the bulge. The presence of hot, young stars near the tips of the outgrowths suggests that the outgrowths have formed relatively recently through processes comprising compression of molecular gas.

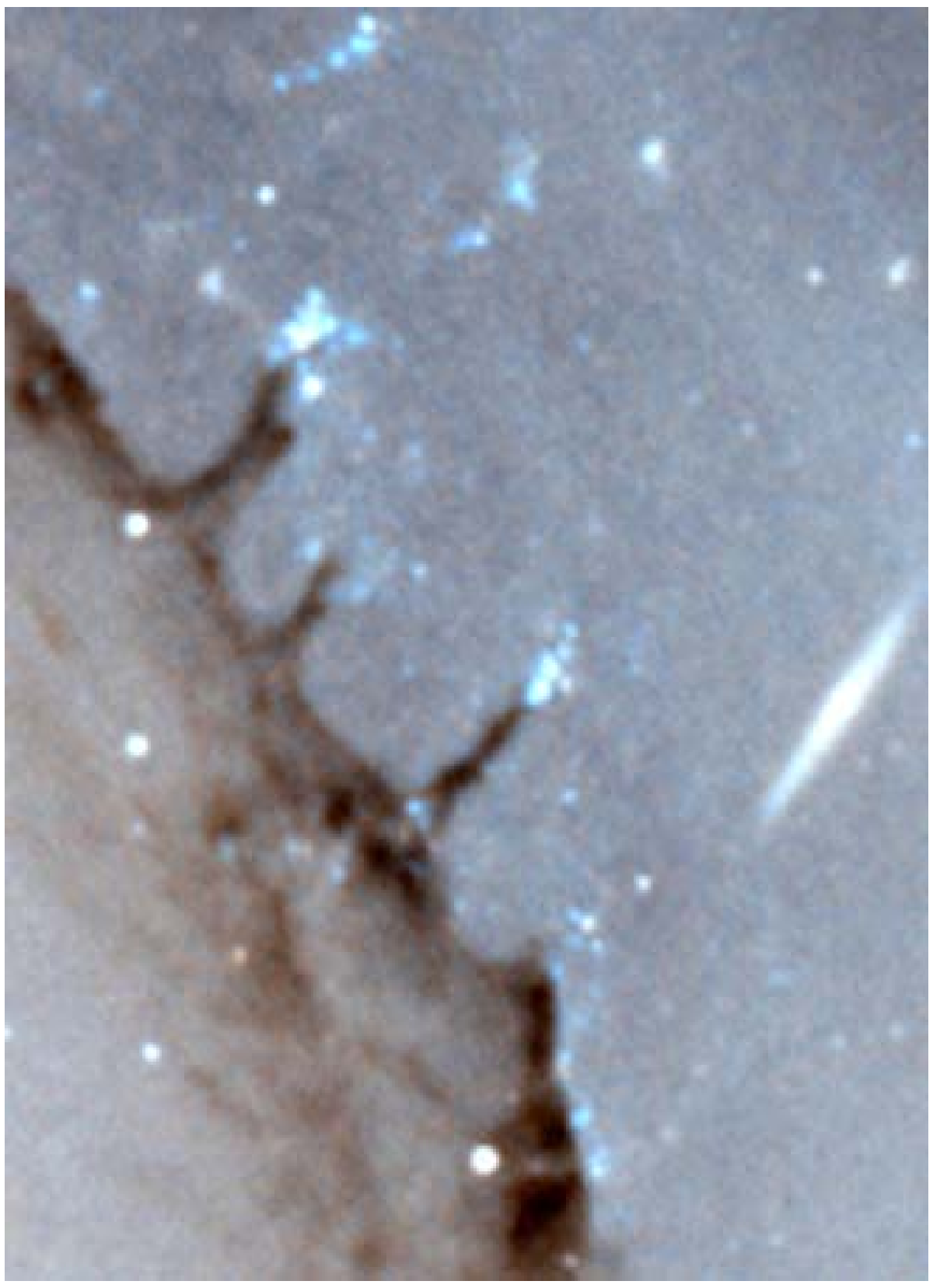

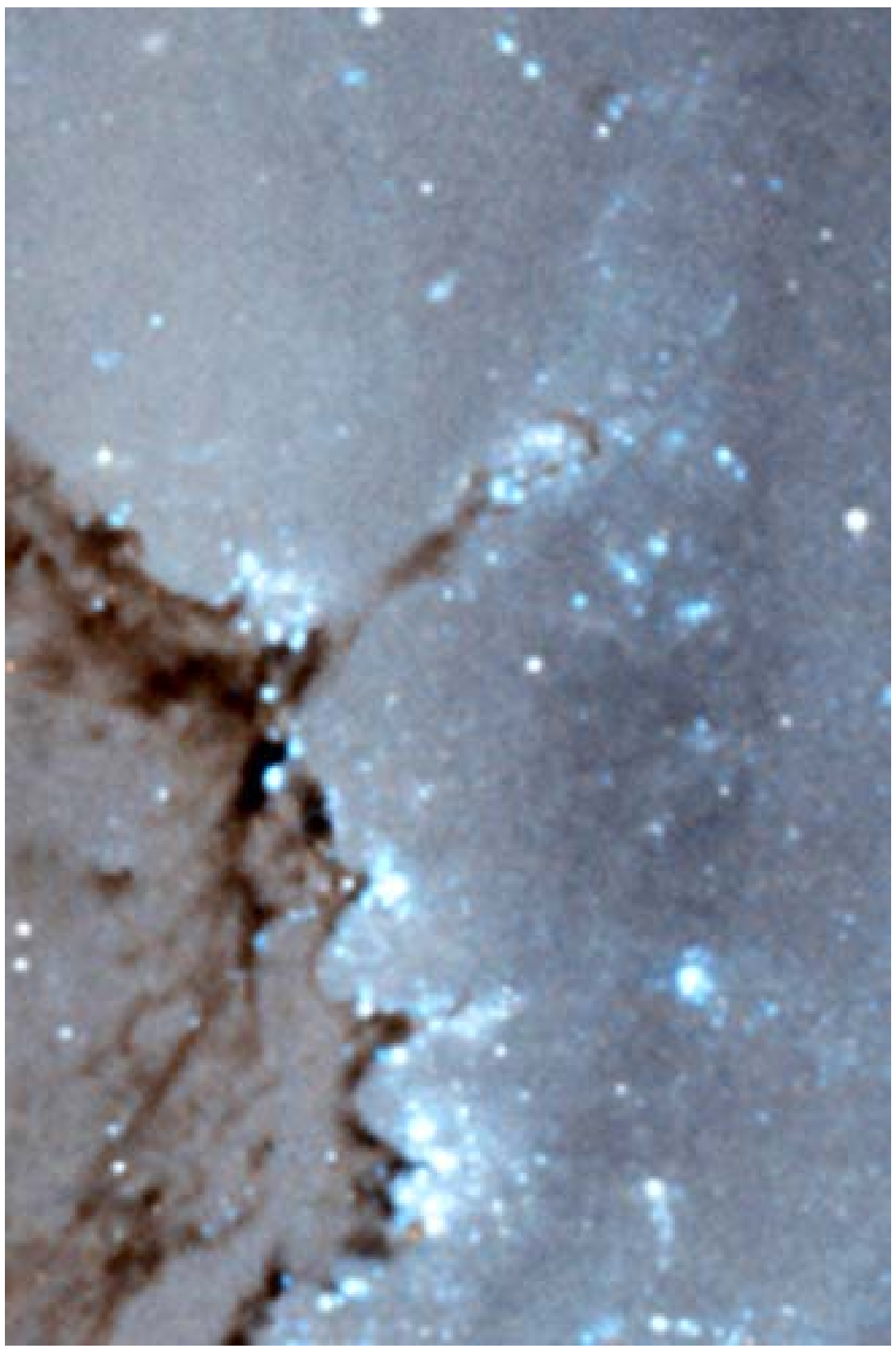

**Figure 3.** (To the left) Some small (~1″ – 2″) outgrowths in the northern part of NGC 4921. The outgrowths protrude nearly perpendicularly outwards from dusty arm filaments. Outside the tips of the outgrowths young, blue stars, or clusters of such stars, are sprinkled. It is to be noticed that the brightest of these objects appear close to the tips of the outgrowths. Image size: 8″ x 12″.

**Figure 4.** (To the right) A number of outgrowths in the northern part of the outer, dusty arm in NGC 4921. A long, filamentary outgrowth showing signs of being twisted is found just above the centre of the image. Two filamentary legs connect it to the arm. In its upper part, the outgrowth is split and forms a structure that may be interpreted as a disturbed loop. In the lower part of the image there are several small, V-shaped outgrowths. Young, blue stars are associated with the outgrowths. Again, the brightest stars are found just outside the tips of the outgrowths. Image size: 11″ x 17″.

About 10″ to the west of the three outgrowths is a longer outgrowth putting out about 6″ (~ 3 kpc) from the outer dusty arm (Figure 4). The shape of this outgrowth indicates that it is composed of twined filaments. Two filamentary legs connect the outgrowth to a dusty arm filament. The head consists of a loop-like structure, partly covered by clusters of young, blue stars. With a slight tilt towards the east, wisps of fainter stars stretch more than 3″ outside the outgrowth. In the lower part of Figure 4, at least five smaller outgrowths put out, all of which are filamentary and associated with clusters of hot stars. Some of these outgrowths have shapes resembling the letter V and the largest one includes a thin, wavy filament pointing outwards from its tip.

An outgrowth of special interest is the long structure seen in the upper half of Figure 2. The outgrowth, denoted #1, mainly consists of two filaments, which in places appear to be intertwined. A mass condensation at the tip to the left constitutes the head of the outgrowth. To the right, the outgrowth is split up into thin legs, which connect it to the outermost dust arm in the lower half of the figure. All in all, the outgrowth ranges more than 35″ (~ 17 kpc). Clusters of young, blue stars are scattered along, and just outside, the structure. In particular, there is a shimmer of blue stars to the left of the head. As is clear from Figure 2, the outgrowth can easily be mixed up with a dust arm similar to the two arms in the lower part of the image. What makes the upper structure to be considered an outgrowth rather than a dust arm is that two or more legs connect it to a dust arm. Later on we shall see that the distinction between the two kinds of object need not be so evident.

When turning to other locations in the galaxy, one finds distinct dust arms, which are similarly equipped with a great number of outgrowths. Especially in the western and southern sectors, the outgrowths are abundant. In the west, several extended outgrowths with lengths of up to ~15″ (~ 7 kpc) can be seen (Figure 5). Some of these (e.g. #2 and #3) may be considered smaller versions of the prominent outgrowth (#1) shown in Figure 2. A notable fact is that the more extended outgrowths bend in the anticlockwise direction just as the arms do. Generally, the tips of the outgrowths consist of a mass condensation forming a dark head. In contrast, the legs are often made up of rather narrow and dilute filaments. In some places, they are even hard to distinguish. Near the heads more or less rich clusters of hot, young stars are mostly present.

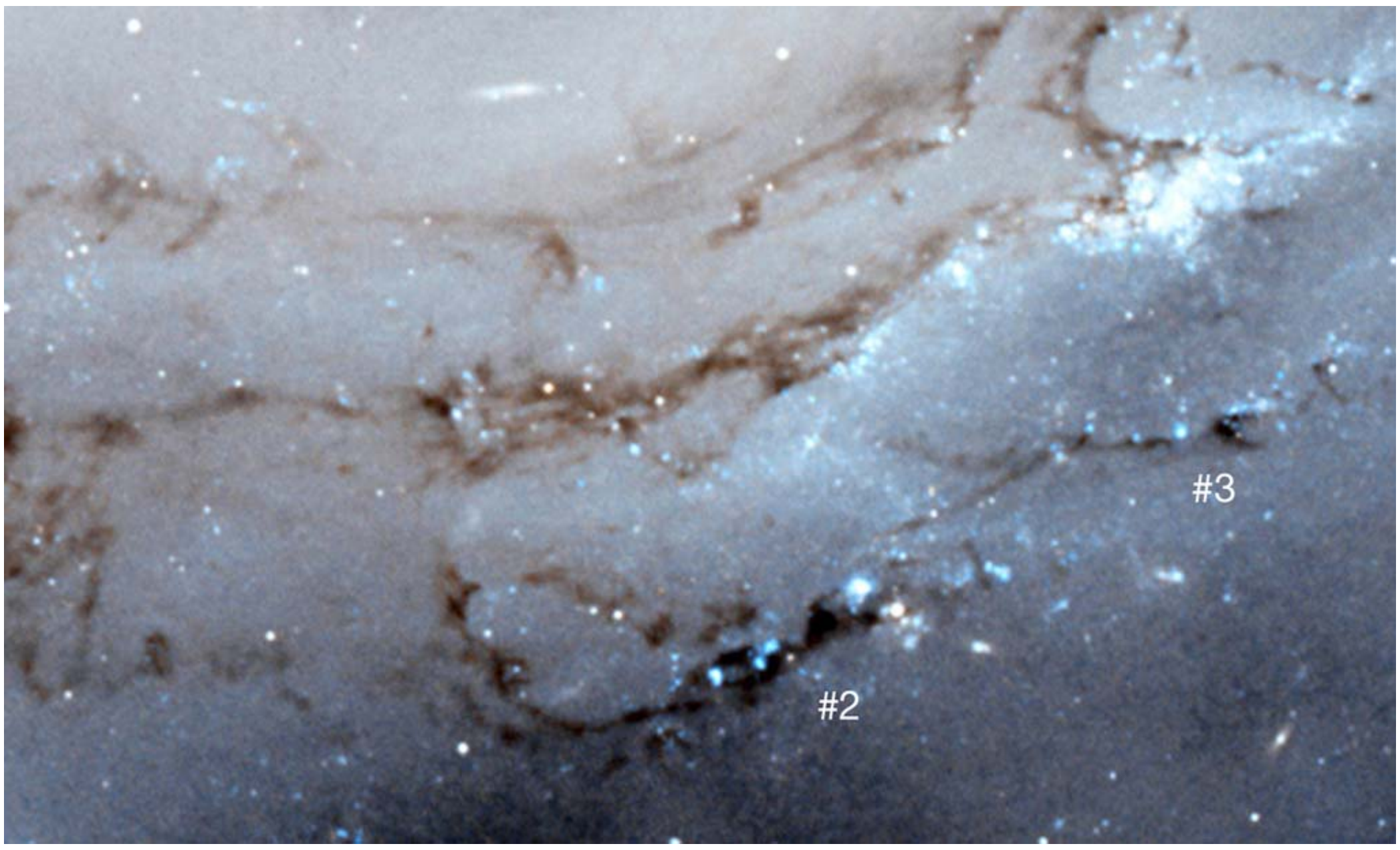

**Figure 5.** A number of outgrowths in the south-western quadrant of NGC 4921. The outgrowths are markedly filamentary and connect to dusty arm filaments. In the upper right-hand corner of the image there are several, partly overlapping, V- and Y-shaped outgrowths, which to some extent are drowned in the glare of luminous, blue stars. Below those a long, bent outgrowth with a claw-like head (#3) is situated. To the left, the outgrowth is split up into two fading legs. Another similar, but more robust outgrowth (# 2) with two legs is found in the lower central part of the image. Image size: 29″ x 17″.

Interestingly, there also exist outgrowths that are oriented inwards in NGC 4921. A number of such outgrowths are to be found in e.g. the southern part of the galaxy. An example of an inwardly oriented outgrowth is shown in Figure 6. Consisting of twisted filaments, the outgrowth has a distinct head with open filaments oriented inwards and a total length of ~10″ (~ 5 kpc). The outgrowth is special in the sense that it originates from a dark, diffuse arm far out in the galaxy. Small clusters of blue stars are associated with its head. There seems to be no counterpart to this solitary structure in the whole galaxy.

Another inwardly oriented outgrowth is present about 20″ south of the centre of the galaxy (Figure 7, *left panel*). Having a length of at least ~ 7″, this outgrowth connects to the innermost arm and points towards north-east. Like many other outgrowths, it appears filamentary and twisted. A peculiarity is that no blue stars are visible near its tip. Instead, hot stars and clusters of hot stars are distributed along part of the arm filament to which the outgrowth is connected. Another inwardly oriented outgrowth is present on the opposite side of the galaxy (Figure 7, *right panel*). Connected to the innermost dust arm in the north-west, this outgrowth is somewhat mottled with slanting sub-structures indicating twists. A condition that distinguishes this outgrowth from the other outgrowths in the galaxy is that its inner tip almost touches the outer region of the central bar. Hence, it cannot be excluded that the outgrowth is under some gravitational influence by the bar. Unlike most of the outwardly that distinguishes this outgrowth from the other outgrowths in the galaxy is that its inner tip almost touches the outer region of the central bar. Hence, it cannot be excluded that the outgrowth is under some gravitational influence by the bar. Unlike most of the outwardly turned outgrowths, the inwardly turned ones shown in Figure 7 have no visible mass condensations near their tips. Hence, it is not surprising to find, that there are no signs of young, blue stars just outside their tips.

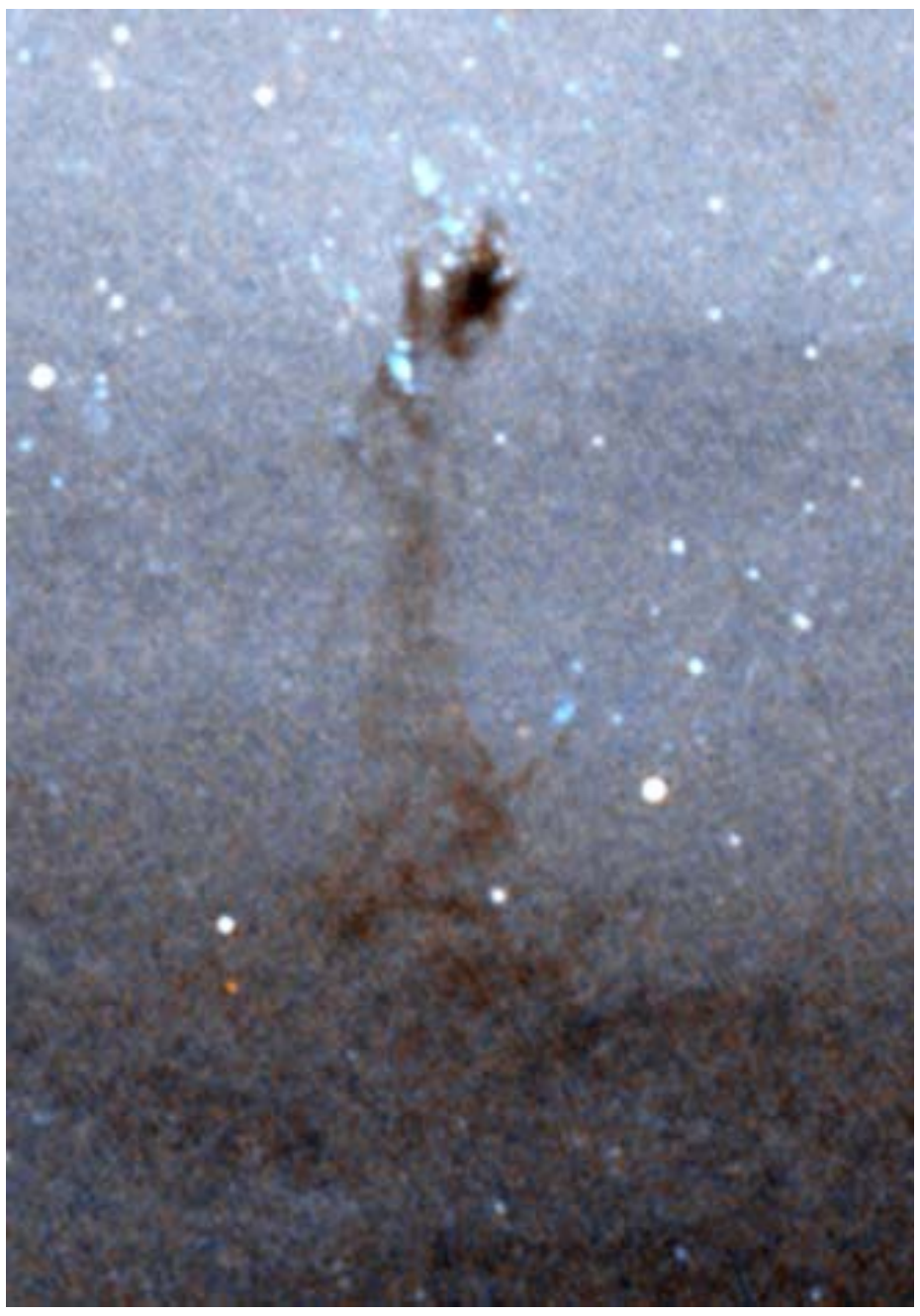

**Figure 6.** An inwardly oriented outgrowth situated ~ 50″ south-west of the nucleus of NGC 4921. This outgrowth is special in the sense that it emanates from a diffuse, dust arm running far out in the galaxy. The arm is nearly horizontally oriented. With a length of ~ 10″ the outgrowth is made up of two or more filaments, which appear to wind around each other. At the tip of the outgrowth is a massive head with a few short, open filaments. Small clusters of young, blue stars can be seen in and just outside the head. Image size: 11″ x 16″.

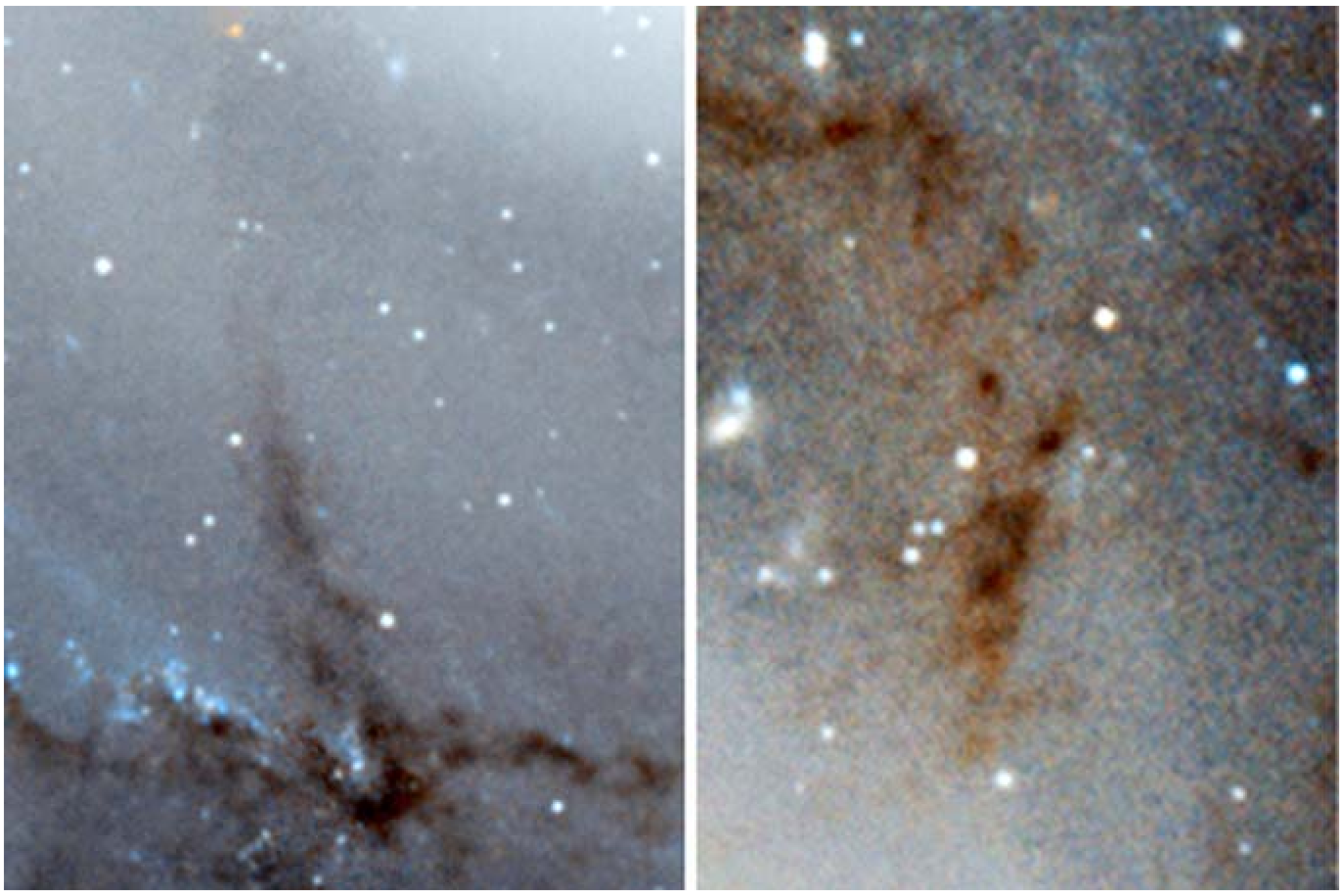

**Figure 7.** Two outgrowths pointing inwards in NGC 4921. The *left panel* shows an outgrowth situated ~ 20″ south-south-west of the nucleus. The outgrowth consists of at least two filaments, which appear to be intertwined. Image size: 11″ x 14″. The *right panel* shows an outgrowth that is situated ~ 15″ north-west of the nucleus. This outgrowth is patchier than the one to the left and presents inclined filamentary pieces. Image size: 8″ x 12″. Both outgrowths are connected to roughly azimuthally oriented arm filaments. No mass condensation is to be seen at the tips of the outgrowths. Neither are there any associations of young, blue stars near their tips.

Most of the outgrowths turn out to be small. In fact, the majority of them are shorter than 3″ while outgrowths exceeding ~12″ must be considered rare objects. This indicates that most of the outgrowths stay small and rarely grow larger than a few arc seconds.

## 3 Morphology of NGC 7049

NGC 7049 is a lenticular, unbarred galaxy (SA(s)$0^0$, e.g. Comerón et al. 2010) with a prominent dust ring. The galaxy is the brightest member of a small cluster in Indus, and may earlier have collided and merged with one or more members of the cluster. As argued by Finkelman et al. (2012) most E/S0 galaxies with dust lanes have acquired their cold dust and molecular gas externally. The distance to NGC 7049 is estimated to about 30 Mpc (Corsini et al., 2003). Like NGC 4921, NGC 7049 includes a variety of dusty arms and filaments as shown by the HST, *ACS* image in Figure 8. However, the number of filaments and arms is here much larger than in NGC 4921. Although present in most of the galaxy, the dusty features are to a considerable degree concentrated to a ring-shaped structure running well outside the centre of the galaxy. The outer diameter of the ring amounts to ~ 47″, corresponding to ~ 7 kpc (1″ ≈ 145 pc). The dusty arms and filaments are most distinct on the near side of the galaxy. On the rear side they are partly swamped with diffuse light emanating from innumerable halo stars. The ratio of the major and minor axes of the ring structure indicates that the plane of the ring forms an angle of ~ 35° with the line of sight. Radial velocity measurements show that the ring structure rotates so that its south-western side recedes faster than its north-eastern side (Corsini et al., 2003).

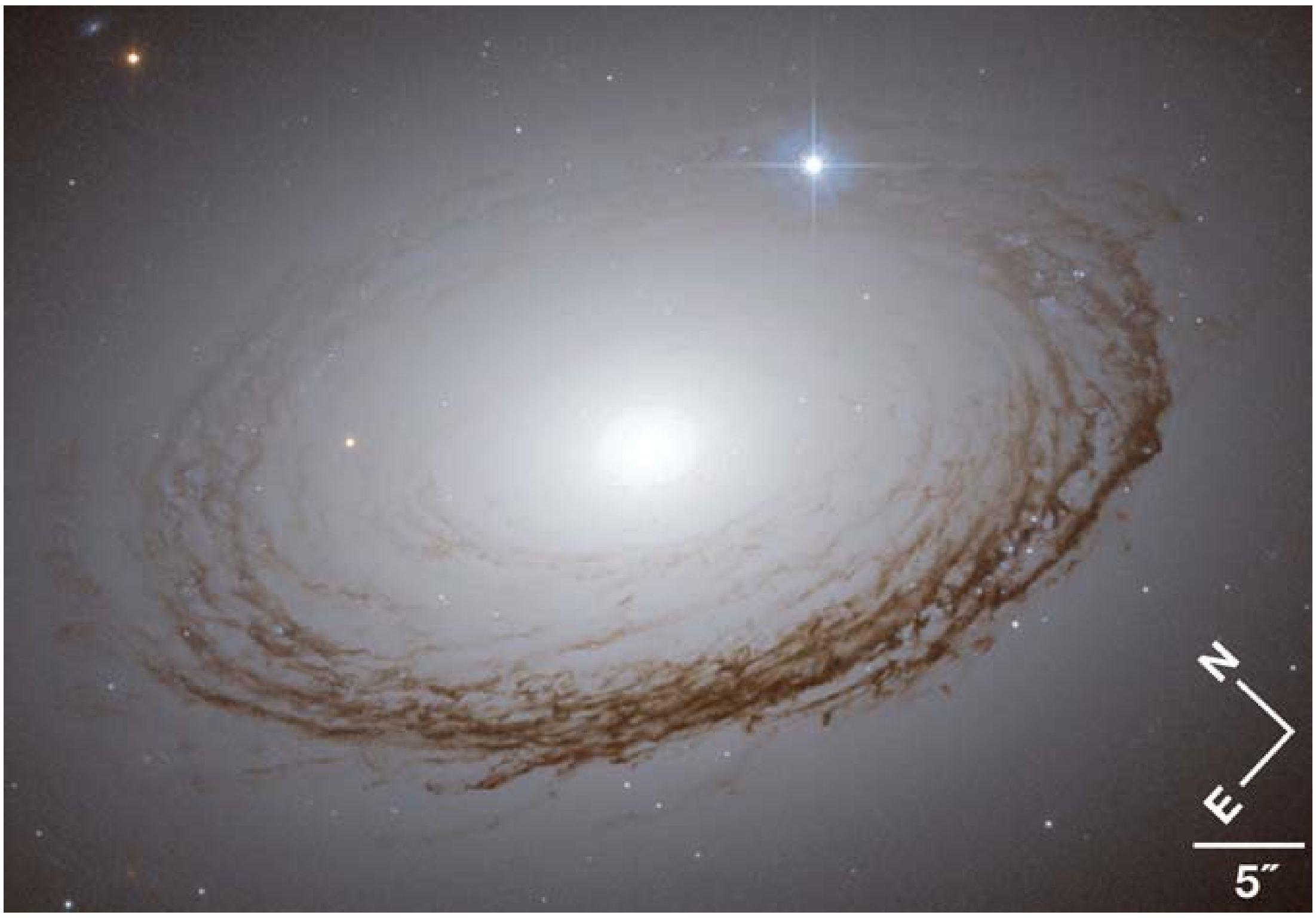

**Figure 8.** An HST, *ACS* image of the giant, spiral galaxy NGC 7049 assembled by F435W and F814W filter images (Credit: NASA, ESA and W. Harris (McMaster University, Ontario, Canada)). The diffuse, bright parts of the galaxy reach well outside the frame of the image. Numerous dust lanes are densely packed within a ring-like structure but there also exist dark features both outside and inside the ring. Of particular interest are the dark outgrowths stretching as well outwards as inwards from the ring structure. The dark features on the far side of the galaxy are partly drowned in the diffuse light of halo stars. The centre of the galaxy is at $21^h\ 19^m.0$, $-48°\ 34'$, (Epoch 2000.0). Also the more detailed images of NGC 7049 shown in Figures 9 to 11 derive from the above image and are oriented in the same way.

In conformity with the situation in NGC 4921, the dust arms in NGC 7049 are furnished with a great number of outgrowths oriented both outwards and inwards in the galaxy. Some of the outgrowths protruding outwards from the dark filaments in the ring structure are made up of small bulges and bends on the filaments while others have lengths amounting to 10″ (~ 1.5 kpc) or more (Figure 8). A closer study reveals that the longer outgrowths are mostly made up of two, thin filaments, which in many places appear to be twisted about one another (e.g. Figures 9, 10). The shorter outgrowths are in general fairly perpendicular to the filaments from which they emerge (e.g. Figure 11). The longer, outwardly oriented outgrowths, on the contrary, are more bent and appear to lag behind their foot points (e.g. #1, #2, #3, Figure 9). In fact, it is hard to tell whether those outgrowths should instead be characterized as dusty, spiral arms. Both shorter and longer outgrowths are composed of at least two dark filaments, often found to be twined around one another. At the tip of some of the outgrowths, a small loop can be seen to connect the two filaments. In other outgrowths again, only a blob is visible. Two legs generally connect the outgrowths to dusty, spiral-arm filaments. The widths of the filaments in the outgrowths typically are of the order 0.″10 – 0.″20 (15 – 30 pc).

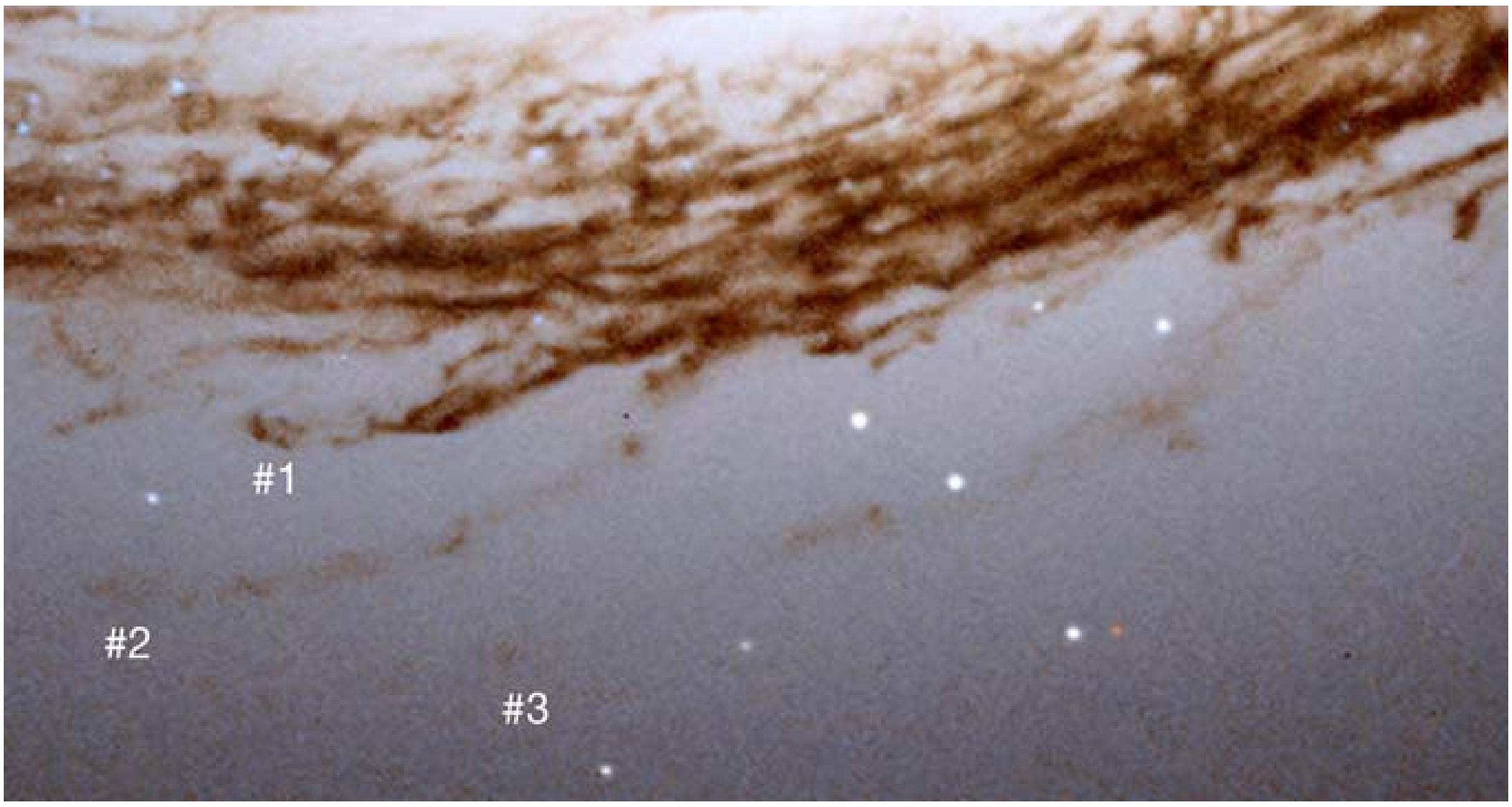

**Figure 9.** A sample of long outgrowths outside the southern part of the ring structure in NGC 7049. Most of the outgrowths seem to consist of at least two dark filaments, which in places appear to be twisted. The outgrowths are tilted so that they are dragging, just as the arms in most spiral galaxies. A loop-like structure can be seen at the end of the outgrowth #1. Image size: 22″ x 12″.

NGC 7049 also contains a number of outgrowths that are oriented inwards in the galaxy. For instance, a few wavy outgrowths with lengths of ~ 10″ (~ 1.5 kpc) may be seen putting inwards from the eastern side of the inner rim of the ring structure (e.g. #4 in Figure 10). Like the outwardly oriented outgrowths, the inwardly oriented ones are connected to dusty arm filaments by means of two legs. Further away from the legs, the outgrowths mostly seem to consist of two intertwined filaments. The longer, inwardly oriented outgrowths are preceding and form deprojected (true) angles of about 30°– 40° with their ring filaments.

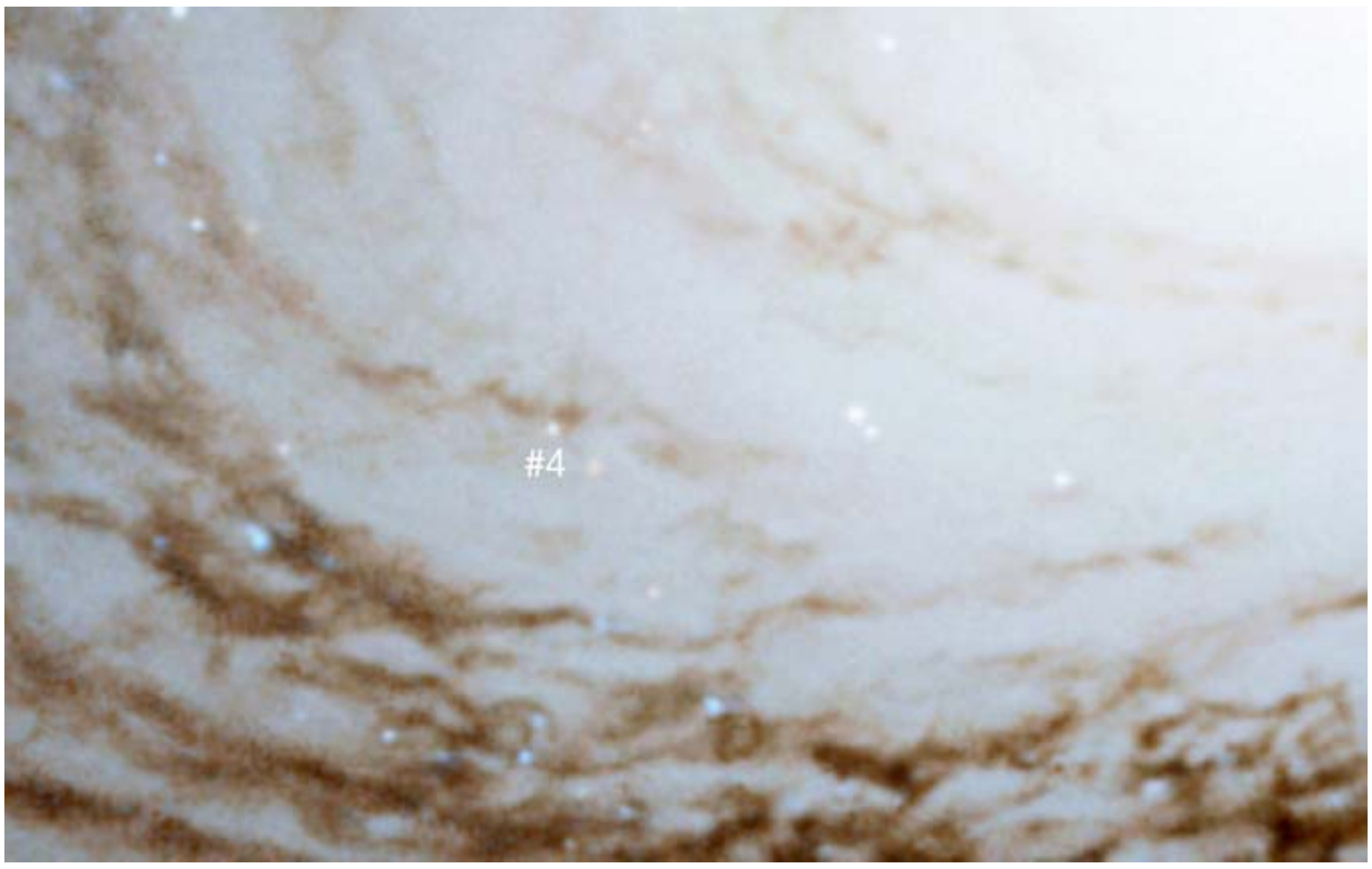

**Figure 10.** A sample of outgrowths stretching inwards from the inner rim of the eastern part of the ring structure in NGC 7049. The outgrowths are made up of two basic filaments, which appear to be wound around one another (e.g. #4). Each of the outgrowths connects to a filament in the ring by means of two legs. Like other inwardly oriented outgrowths, the outgrowths shown are preceding with deprojected tilts of ~ 30°– 40° relative to the ring. Image size: 16″ x 10″.

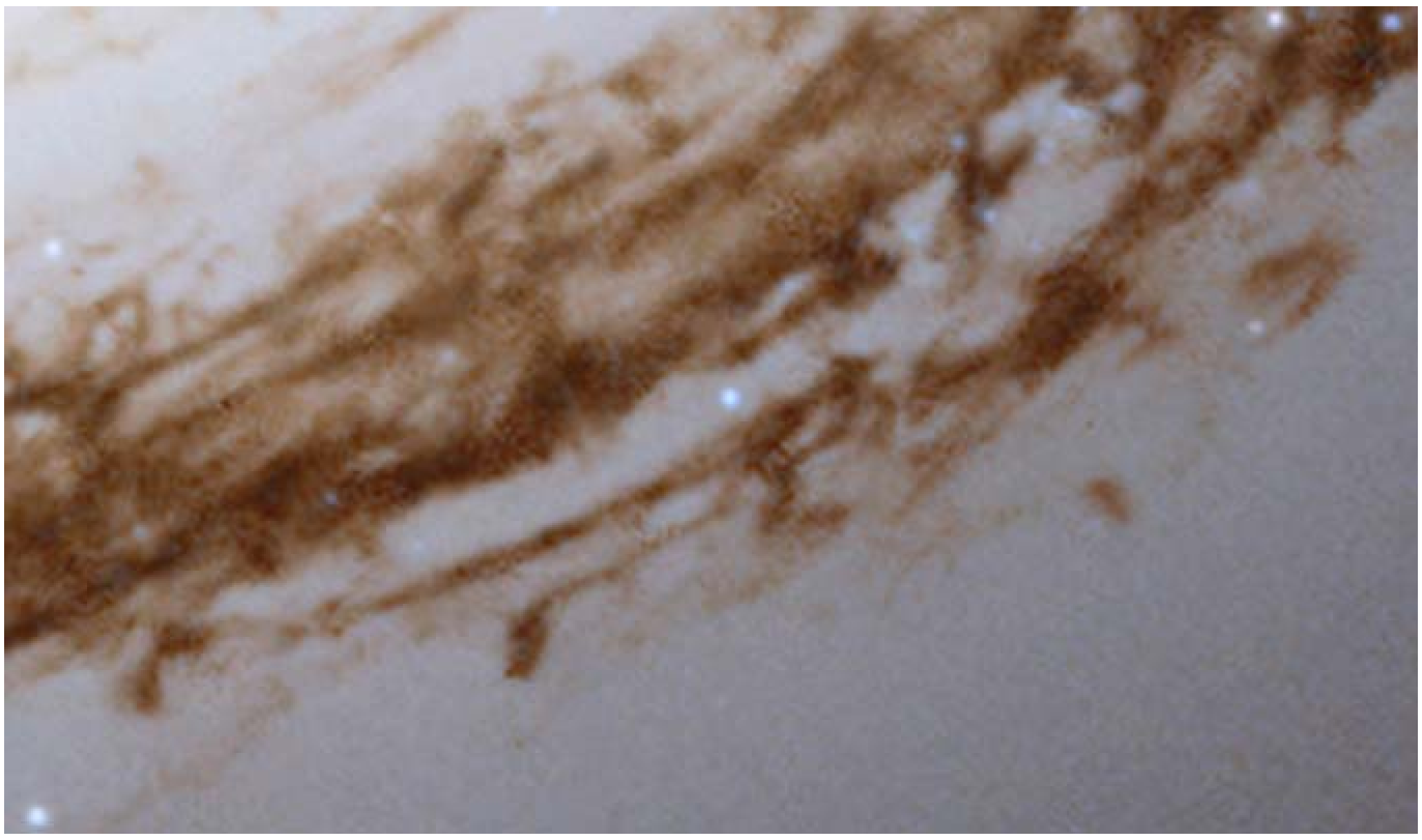

**Figure 11.** A number of small outgrowths (≤ 0.″7) putting out from outer, dusty filaments in the ring shaped structure in the southern part of NGC 7049. Some of the outgrowths show signs of being twisted one or a few turns. It should be noticed, that the short outgrowths are fairly perpendicular to the filaments from which they emerge. Image size: 13″ x 8″.

One may ask oneself whether outgrowths like those in Figure 10, above judged to be inwardly oriented, could in fact turn out to be oriented in the opposite direction? Such outwardly oriented outgrowths would then have similar tilts as the longer, outwardly oriented outgrowths outside the ring structure (Figure 9). What speaks against such an interpretation, however, is the fact that the outgrowths in Figure 10 have legs, which connect them to dark filaments situated near the inner rim of the ring structure. In addition to that, the outgrowths are found to stretch inwards only a limited distance from the inner rim of the ring. We therefore conclude that these outgrowths are indeed oriented inwards in the galaxy.

In accordance with NGC 4921, there are a number of young, blue stars present in NGC 7049. As opposed to the situation in NGC 4921, only few blue OB stars or clusters of such stars can be seen near the protruding structures outside the ring structure in NGC 7049. Instead, the hot stars are generally scattered among the dusty arms and filaments inside the ring structure.

Above, we have used the terms *filament* and *filamentary* to describe the narrow and elongated lanes, which make up the outgrowths and dusty arms. This immediately carries the thoughts to long, tubular formations. But the question is: could the narrow, elongated features be of other geometries? For instance, could the features be interpreted as sheets in the form of shocks or shells seen edge-on? We do not think so. Several arguments strongly speak in favour of that they really are filamentary (Carlqvist 2010). A main argument is that it would be extremely unlikely if all such surfaces would be oriented precisely towards us. The twining of the structures constitutes a further argument.

Summarizing the studies of the outgrowths in NGC 4921 and 7049, we find that, from a morphological point of view, many of the outgrowths are very similar to the mammoth trunks in NGC 1316 and elephant trunks in H II regions. Like the mammoth and elephant trunks, the outgrowths are markedly filamentary and often appear to be twined. For those outgrowths in which the twining is possible to observe, the helix angle is in the mean found to be ~ 45° with a scattering of about 10°. If instead of the morphology, the orientations of the three kinds of structure are considered, there are clear differences. The outgrowths are oriented both outwards and inwards in the two galaxies studied whereas elephant and mammoth trunks are oriented only inwards. This difference may be of some significance, since it may offer a clue to how the outgrowths once were formed.

Another point, which could have some bearing on how the outgrowths were formed and developed, concerns the relationship between the outgrowths and the dusty spiral arms. In the previous section it was found hard to distinguish the prominent outgrowth #1 in NGC 4921 from a dusty spiral arm. Still, it turns out to be even harder to tell whether some of the longer, outwardly oriented structures outside the ring structure in NGC 7049 are to be classified as spiral arms or just outgrowths. So the crucial question is: outgrowths or spiral arms? The presence of connecting legs points to the former identity whereas the overall impression favours the latter interpretation.

To illuminate these problems, we shall aim at finding out how the outgrowths once were formed. In this matter we will seek assistance from a different source. Since many of the dark outgrowths in NGC 4921 and 7049 are in several respects so similar to elephant trunks, it is natural to investigate whether the mechanism behind the outgrowths could have some elements in common with the formation mechanism of elephant trunks. Hence, a high priority task is to have a closer look at elephant trunks.

## 4 Elephant trunks

### *4.1 Properties and theories of elephant trunks*

Elephant trunks are elongated, dusty clouds mostly seen in silhouette against bright H II regions (Minkowski 1949; Bok et al. 1949). Generally, trunks are found in close association with molecular clouds where one or more luminous OB stars have recently been formed. Winds and radiation emanating from the OB stars blow up an expanding cavity in the cloud. From the border between the inner, hot and relatively dilute H II region and the outer, cold and dense molecular cloud, elephant trunks put inwards like fingers in the cavity. With the access to more detailed images captured by modern telescopes, it has been increasingly clear that elephant trunks are to a high degree filamentary (Carlqvist, Gahm & Kristen 2002, 2003, hereafter CGK02 and CGK03). Mostly, the trunks are shaped either as the letter V or as the letter Y with the point or stem, respectively, roughly pointing towards the OB stars. In particular, when the stem of the Y is somewhat transparent, intertwined filaments are often seen in it. A good example of a trunk with such features is displayed in Figure 12. Another fine illustration of a Y-shaped and twisted elephant trunk is found in Figure 1 in CGK03.

A few years after Minkowski and Bok et al. had drawn attention to the phenomenon of elephant trunks Spitzer (1954) and Frieman (1954) proposed a theory of trunks based on the Rayleigh-Taylor instability. In that theory trunks are supposed to be formed when dense gas in the original molecular cloud is pushed ahead by thinner and hotter gas in the cavity surrounding the OB stars. Later, new ways of approaching the trunk problem were advanced (Osterbrock 1957; Pikel′ner 1973; Pikel′ner & Sorochenko 1974; Schneps, Ho & Barrett 1980). Although appealing, all these theories turned out to have difficulties in explaining the basic, filamentary character of elephant trunks as well as their twisted geometry. In consequence of those drawbacks, a new theory of elephant trunks was worked out some years ago (CGK02, CGK03). Partly based on magnetized filaments and magnetohydrodynamics (MHD), the theory is, among other things, capable of accounting for both the filamentary nature and twined structure of trunks. A more detailed account of this filamentary theory of elephant trunks is given in Section 4.2.

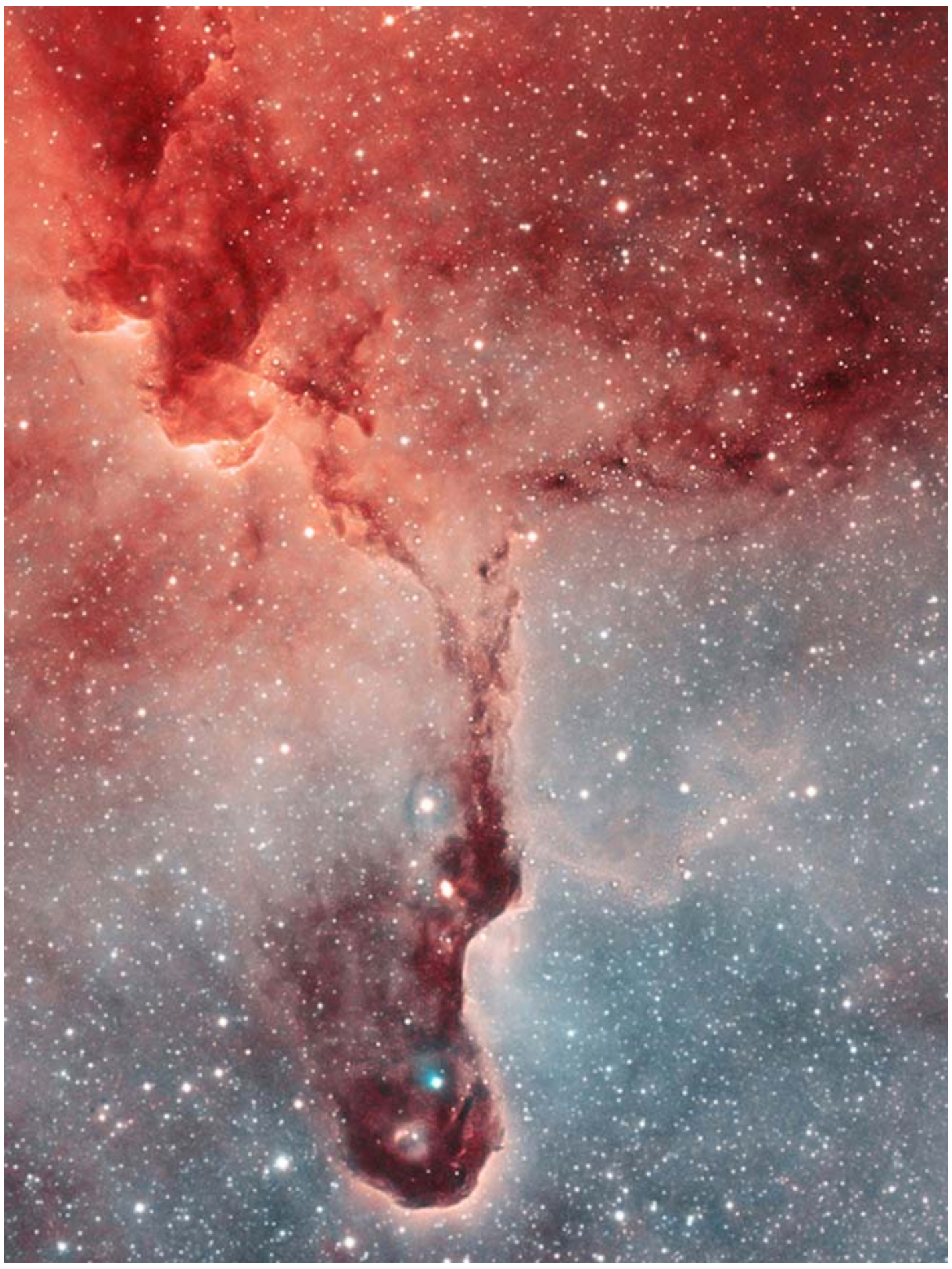

**Figure 12.** One of the larger elephant trunks in the galactic H II region IC 1396. The trunk is a good illustration of a Y-shaped trunk consisting of two legs, a stem, and a head. The stem of the Y shows clear signs of being twisted and so do the two legs. The length of the stem is ~ 17′ corresponding to ~ 4 pc in projected length at an estimated distance of ~ 750 pc. The head is situated at $21^h\ 36^m.9$, + 57° 31′, (Epoch 2000.0). The colour picture is a composite of images taken in the lines of Hα, O III, and S II. Image size: 27′ x 36′. For a full version of the image, see Apod, November 6, 2010. Credit & Copyright: R. Geissinger, Remseck, Germany.

Efforts to find out the mechanisms behind elephant trunks have in recent time also been made along quite different lines of approach. In particular, the ionization front separating the H II region and the surrounding molecular cloud has attracted special interest. Thus, Redman et al. (1998), Williams, Dyson & Hartquist (2000) have in some detail clarified the jump conditions of magnetic ionization fronts by means of the continuity equations. Furthermore, Williams & Dyson (2001) have presented calculations of the internal structures of stationary fronts with oblique upstream magnetic fields.

Great progress in the study of irradiated ionization fronts has also been made by means of various kinds of numerical simulations. Using MHD simulations, Krumholz, Stone & Gardiner (2007) have investigated the global expansion of an H II region in an initially uniform gas threaded by a homogeneous magnetic field. Among other things, they found that the expansion of the H II region perpendicular to the field is suppressed, especially at later stages of the development. More details in the neighbourhood of the ionization front appear in a radiation-MHD simulation starting with a turbulent, magnetized molecular cloud (Arthur et al. 2011). As a result of the influence of the radiation, a few pillar-like structures appear in close association to the front while putting inwards in the ionized region. To be noted here is, that also in the absence of a magnetic field a similar, but purely hydrodynamic, simulation results in roughly the same structures.

In this connection, it is appropriate to mention that also other radiation-hydrodynamic simulations with diverse initial conditions show that compressed structures can arise adjacent to the H II region, some of which being pillar-like (Melemma et al. 2006; Lora, Raga & Esquivel 2009; Gritschnader et al. 2010; Ercolano et al. 2012). The investigations by Lora, Raga & Esquivel and Gritschnader et al. comprise plane-parallel geometries, the latter with initial turbulence and the former with no initial velocities but with density fluctuations.

It is also of some interest to note that a number of radiation-simulations starting with smaller or larger clumps of gas have been performed (Williams 2007; Raga et al. 2009; Henney et al. 2009; Mackey & Lim 2011; Bisbas et al. 2011; Gendelev & Krumholz 2012). The simulations, which result in various kinds of geometries for the compressed gas, including pillars, mainly aim at studying the formation of globules and stars. It is concluded that a strong, initial magnetic field can alter the non-magnetized dynamics. In the study performed by Gendelev & Krumholz a magnetized gas cloud is irradiated from its edge, which can result in a very efficient energy injection into the cloud. In conclusion, it should be remarked that there is a clear trend of the numerical simulations over the years implying that the simulations have been increasingly detailed and refined. Obviously, the success of a simulation is closely associated with the skill to foresee the essential, initial conditions in molecular clouds as well as to have the computational capacity to realize them.

### *4.2 The filamentary theory of elephant trunks*

An important element of the filamentary theory of elephant trunks is a magnetized filament. In the last few decades, such filaments have turned out to be common components of molecular clouds (e.g. Bally et al. 1987; Wiseman & Ho 1996; Nagahama et al. 1998). Moreover, the clouds have been found to be penetrated by strong magnetic fields capable of influencing the dynamics of the clouds (see Section 5.1). Since elephant trunks are known to be formed out of molecular clouds, it is natural to expect that also the trunks must to some extent be governed by strong magnetic fields. Within the filament, the magnetic field is thought to run along the filament while spiralling around its axes. Often the filaments consist of a number of sub-filaments, which wind around each other and form a *magnetic rope* – a term coined by Babcock (1961) to describe magnetic flux tubes in the solar atmosphere. The reason for the twisting may be sought in turbulent motions known to be present in molecular clouds.

According to the theory proposed, the expanding cavity around the OB stars gradually sweeps up dusty molecular gas into a shell. Many filaments embedded in the gas are also swept up in this process. Only if there is a local mass condensation in a filament, it may lag behind the shell due to its larger inertia. Since the magnetic field is intrinsically divergence-free (div $\boldsymbol{B} = 0$), and in addition well frozen-in to the gas in the filament (see Section 5.1), the filament must form a continuous unit. As a consequence of that, any mass condensation lagging behind the expanding shell must be connected to the rest of the filament in the shell by means of two filamentary legs. The filamentary magnetic field also influences the filament in other ways. For instance, it protects the filaments from being quickly eroded by the hot gas in the H II region. Another effect, especially important to the theory, is that the field can assist in the shaping of the filament.

If the magnetic field inside the filament is not twisted too much, the combined action of the expansion of the shell and the enhanced inertia of the mass condensation will force the filament to assume the shape of the letter V. The point of the V containing the mass condensation is oriented inwards from the shell, towards the luminous OB stars near the centre of the H II region. If, on the contrary, the twisting of the magnetized filament exceeds a certain, critical limit, the filament just above the point of the V is wound up into a double helix. The filamentary part stretching inward from the shell now more resembles the letter Y. The stem of the Y is constituted by the double helix pointing towards the luminous OB stars. The theory thus offers an account of both the V-shape and the Y-shape found with elephant trunks.

The various stages of the development of the double helix can be described by means of two principles. The first principle implies that the system tends to minimize its total energy. The second principle tells that the helicity of the magnetic field in the filament

$$H = \langle \boldsymbol{A} \cdot \boldsymbol{B} \rangle \qquad (1)$$

should be constant (Woltjer 1958; Moffat 1969). $\boldsymbol{A}$ is here the vector potential for the filamentary magnetic field $\boldsymbol{B}$ while $\langle\ \rangle$ stands for a volume average over the filament. For a more thorough description of the theory behind the double helix mechanism, see CGK03 and references therein.

The filamentary theory of elephant trunks including the double helix mechanism can be simply illustrated by means of a mechanical analogy model (CGK02). The model is founded on the well-known fact that magnetic field lines can in many respects, even where MHD phenomena are concerned, be compared to elastic strings (Faraday 1839–1855; Alfvén 1950a). The model consists of a bundle of elastic strings, firmly attached to the ends of a horizontally oriented bar, with a weight hanging down from the middle of the bundle. The bundle of strings represents the filament with its frozen-in magnetic field lines while the gravitational force acting on the weight simulates the inertia force of the mass condensation in the filament. The attachment of the bundle to the bar imitates the connection of the hanging-down part of the filament to the rest of the filament in the shell. If now, the bundle of strings is only slightly twisted, or not twisted at all, the bundle hangs down in the form of a V with the weight at its point (Figure 13, *left panel*). If instead the bundle is twisted beyond a certain critical limit, it is wound up into a double helix just above the point of the V so that the overall shape resembles a Y (Figure 13, *right panel*).

The analogy model is also capable of illustrating motions in the double helix. If the bundle of strings is twisted well beyond the critical limit, new turns are added to the double helix leading to a prolongation of the double helix. At the same time, the double helix must rotate about its axis. In this connection, it is worth mentioning that such a rotation has been observed in elephant trunks. Observations of four different trunks in the $^{12}$CO and $^{13}$CO lines show that all four trunks rotate (Gahm et al. 2006). The periods of rotation found are all of the order of one to a few Myr. Also, observations by Schneps, Ho & Barrett (1980) lend support to these findings.

Summing up the most significant properties of the filamentary model of elephant trunks one finds that the model can account for several prominent features observed with trunks. First, i) the basic element of the model, the magnetized filament, is an almost necessary constituent since such filaments generally make up a considerable fraction of molecular clouds, out of which elephant trunks are formed. Furthermore, ii) the model can naturally account for both the Y- and V-shapes observed with trunks as well as for iii) their filamentary and twisted structures. These structures are most evident in fluffy trunks but can also be detected in denser trunks as tilted bands, which stand out against the trunks. Another property of the model is that iv) it naturally includes magnetic fields capable of influencing the trunks. Finally, v) the model is able to give a description of the rotations observed in some trunks.

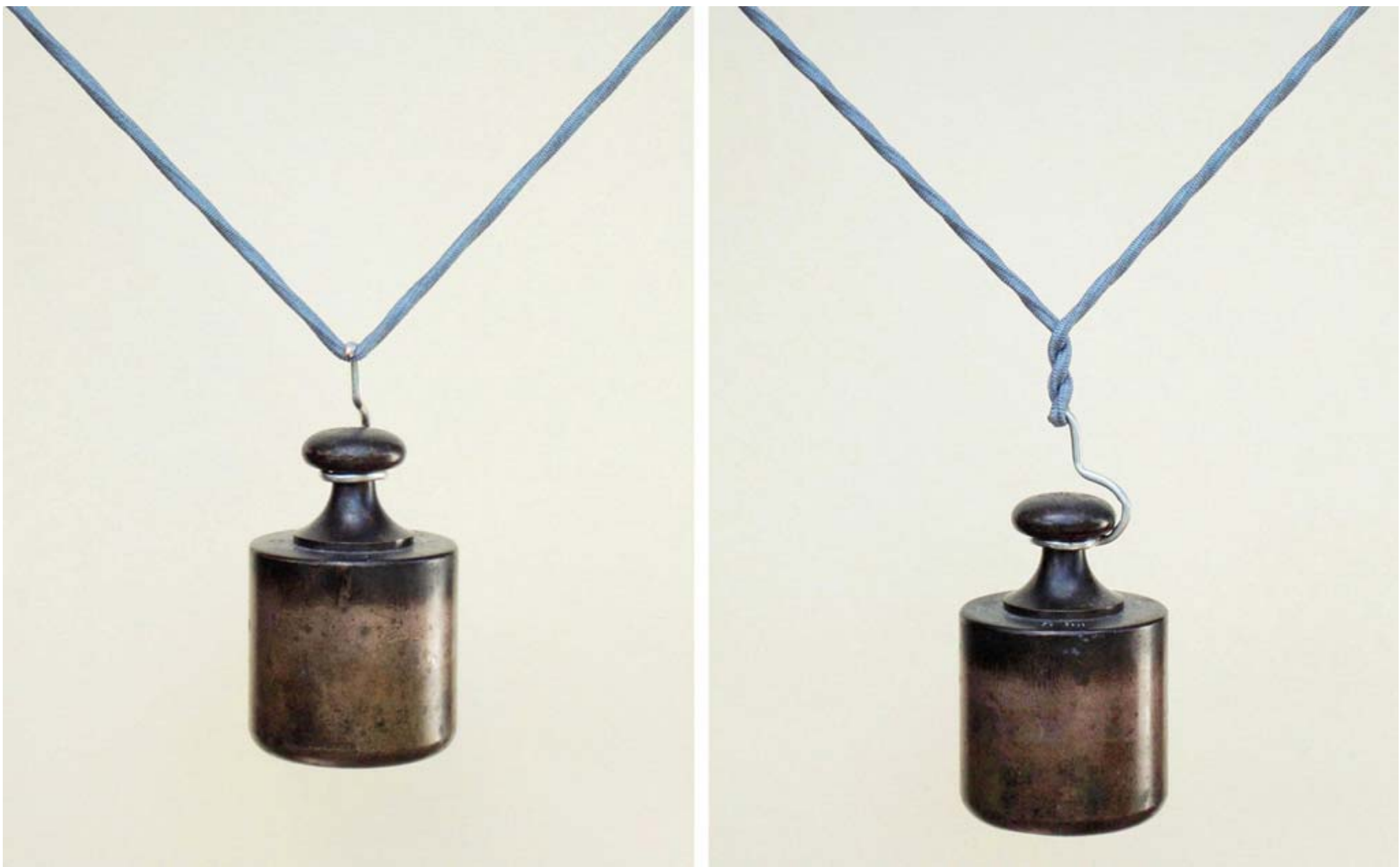

**Figure 13.** Mechanical analogy model illustrating the double helix mechanism. The model consists of a weight hanging down from the middle of a bundle of elastic strings fastened at the ends of a horizontal bar (not visible on the image). The elastic strings represent the magnetic field lines while the force exerted by the weight simulates the inertia force of a mass condensation in the filament. If the bundle of strings is only moderately twisted, or not twisted at all, the bundle assumes the shape of the letter V (*left panel*). If the bundle of strings is more twisted so that the twist exceeds a certain critical limit, the bundle forms a double helix next to the weight (*right panel*). The bundle now has an over-all shape resembling the letter Y.

## 5 Model of the outgrowths

### *5.1 Dusty arm filaments*

An important prerequisite of the outgrowths is the dusty arm filament out of which the outgrowths protrude. In the search of a model of the outgrowths it is therefore suitable to begin with considering the arm filament. There are plenty of such filaments in the dusty spiral arms of NGC 4921 and 7049. A glance at the HST images of the two galaxies studied quickly reveals that the dark filaments in the spiral arms are very long and narrow. Their lengths can amount to tens of thousands of parsecs while the typical widths are ~ 0.″10 – 0.″25 (~ 50 – 120 pc) in NGC 4921 and ~ 0.″10 – 0.″40 (~ 15 – 60 pc) in NGC 7049. It is to be noticed, that the lower limits found are close to the resolving power of the HST suggesting that even narrower filaments may exist. The fact that the filaments form long and narrow structures strongly indicates that they are kept together by magnetic fields.

Owing to the fact that NGC 4921 and 7049 are fairly distant, the physical conditions in the dark filaments in the two galaxies are not known in any detail. A possible way to get away from this difficulty and learn more about the properties of the filaments is to benefit from what is known about corresponding dark clouds in our own Milky Way Galaxy. Here, the dark arms and filaments are made up of dusty, molecular gas with temperatures of $T_g \approx 10 - 30$ K. The gas densities of such clouds typically are $n_g \approx 10^8 - 10^9$ m$^{-3}$ = $10^2 - 10^3$ cm$^{-3}$ although both higher and lower densities occur locally. The clouds mainly consist of $H_2$ molecules with a smaller share of He atoms. About 1% of the cloud mass is constituted by dust particles of sub-micron sizes. In addition to that, there also exists a tiny fraction of ions and electrons amounting to ~ $10^{-7} – 10^{-4}$ (Hjalmarsson & Friberg 1988; Black & van Dishoeck 1991). Although little in relative number, the charged particles are of decisive importance for how the gas behaves. The presence of ions and electrons transforms the gas into a medium fulfilling all the demands for a *plasma* (e.g. Chen 1984). Another important property of the molecular-cloud plasma is that it is represented by a very large magnetic Reynolds number ($R_m$ >> 1) implying that the magnetic field lines are well frozen-in to the plasma for long periods of time (e.g. Alfvén & Fälthammar 1963; Spitzer 1978; Ruzmaikin, Shukurov & Sokoloff 1988). Even though the molecular-cloud medium is thus strictly made up of plasma, we shall henceforth stick to the convention and call it a gas, although well aware of its special properties.

Measurements of the polarization of light from stars situated beyond elongated, molecular clouds in the Milky Way Galaxy show that the clouds are penetrated by comparatively strong magnetic fields having a significant component along the long axis of the clouds. For several decades it has been clear that the strength of such magnetic fields $B$ is closely coupled with the density of the molecular gas $n_g$ (Troland & Heiles 1986; Myers & Goodman 1988; Valleé 2003). Disregarding some scattering, the mean strength of the magnetic field is well described by the relation

$$B \approx C_1 n_g^{1/2} \quad , \tag{2}$$

where $C_1$ ≈ 1.6 x$10^{-13}$ (SI-units). Hence, for typical gas densities of $n_g \approx 10^8 – 10^9$ $m^{-3}$ = $10^2$ – $10^3$ $cm^{-3}$ the strengths of the magnetic field are of the order $B$ ≈ 2 – 5 nT = 20 – 50 μG, respectively.

Using equation (2), we can express the magnetic energy density (or magnetic pressure) in molecular clouds as

$$p_m = \frac{B^2}{2\mu_0} \approx \frac{C_1^2}{2\mu_0 k T_g} p_{th} \quad . \tag{3}$$

Here, $\mu_0$ is the permeability of free space, $k$ is Boltzmann´s constant, while $p_{th} = k n_g T_g$ is the thermal energy density (or thermal pressure) of the gas. Putting $T_g$ = 15 K, we obtain $p_m / p_{th} \approx 50$, which shows that the magnetic energy density largely dominates the thermal energy density. If the turbulent energy density of the gas is also taken into account, the total kinetic energy density of the gas may in some places rival the magnetic energy density while in other places again magnetic energy density still dominates. All in all, this means that the magnetic field is of great significance and must play an important role for the dynamics of the molecular gas.

The discussion above certainly deals with molecular clouds in the Milky Way Galaxy but there is little reason to believe that the physical conditions in the dusty arms of NGC 4921 and 7049 should be very different. In all essentials, the properties of the galactic molecular clouds should therefore be applicable also to the dusty clouds in those galaxies. As pointed out in Sections 2 and 3, the dusty spiral arms in NGC 4921 and 7049 generally consist of two or more dark filaments. By analogy with the dust arms in the Milky Way Galaxy, relatively strong magnetic fields are expected to run along also the dark arm filaments in these galaxies. The magnetic field lines may be likened to reinforcement bars, which exert a stabilizing effect on the filaments and prevent them from being torn into pieces when exposed to strains. In addition to that, the presence of the field also considerably reduces evaporation of gas from the filaments.

### *5.2 Formation mechanism of the outgrowths*

When attempting to work out a new model of a specific object, it is suitable first to pay attention to the most basic properties of the object. As regards a prospective model of the outgrowths, such basic properties that should first be taken into consideration are the filamentary and twisted structures of the outgrowths. Another crucial question to tackle is of course how an outgrowth can at all start to grow out from a dusty arm filament. One may note, that similar questions were central also at the accomplishment of the filamentary theory of elephant trunks several years ago (Section 4.2). The solution of the problems at that time was managed by the double helix mechanism. It is therefore reasonable to search for a model of the outgrowths in a similar direction.

A natural way of approaching the problem is to try a tool developed by Hannes Alfvén long ago (Alfvén 1950b; Alfvén & Fälthammar 1963). Alfvén considered a long cylindrical filament consisting of a well-conducting, incompressible fluid. The fluid is penetrated by an initially homogeneous and axis-parallel magnetic field, which is supposed to be well frozen-in to the fluid. If now, one of the ends of the filament is turned some angle around the filamentary axis while the other end is kept fixed, the field lines in the filament are twisted into spirals. Alfvén showed that if the twisting exceeds a certain critical limit, the filament must locally form a loop. The critical limit for loop formation was found to be reached when the azimuthal component of the magnetic field in the outer parts of the filament becomes of about the same magnitude as the axial component. Since the loop represents the first stage at the formation of a double helix with a closed top, one might expect that this process, if continued, should lead to a double helix. Unfortunately, this turns out not to be the case. Alfvén´s loop mechanism does not lead all the way to a double helix. If the filament is further twisted, well beyond the critical limit, only a series of loops, situated side by side, are formed on the filament (Alfvén 1950b). (Actually, Alfvén worked out his mechanism for other purposes than those we are concerned with here.) Even if Alfvén´s mechanism thus does not lead to a fully developed double helix, it still points out a general course towards the goal.

According to the filamentary theory of elephant trunks, a double helix is formed if an additional force acts on the magnetized filament at the same time as the filament is twisted beyond a critical limit. As pointed out in Section 4.2, the force in that theory is caused by the expansion of an H II region coupled with the enhanced inertia of a mass condensation in a filament. Remarkably enough, a similar situation occasionally occurs in the dusty arms of spiral galaxies. Observations show that expanding H II regions, mainly energized by clusters and associations of luminous OB stars and subsequent supernova explosions, now and then develop in the dusty arms of spiral galaxies. The expansion causes the dark arm filaments to be pushed aside so that the dusty spiral arm is locally widened (Figure 14). The environment here is rather much suggestive of the environment in which elephant trunks are formed. If any of the dark arm filaments contains a local mass condensation – in Figure 14 represented by a blob – it will lag behind the rest of the filament because of its larger inertia. Being held together by its intrinsic magnetic field, the filament must then take on a V-shape. If, in addition, the filament is sufficiently twisted, part of the filament must form a double helix next to the point of the V. Thus, a Y-shaped filament is produced in agreement with the theory of elephant trunks. Using a simplified picture, one may say that the additional force provides a point of preference, or seed, where all the loops in Alfvén´s scheme are concentrated into a double helix. A more strict derivation can also in this case be obtained by means of the two principles discussed in Section 4.2. Of course, it is far from certain that any of the dark filaments that are pushed outwards by the expanding H II region both contains a large mass condensation and is sufficiently twisted. Probably only a fraction of the more powerful H II regions formed in the dusty arms are capable of producing an outgrowth containing a double helix. From the scheme depicted, it is clear that the V-shape or Y-shape formed, can point in any direction in a plane perpendicular to the dust arm.

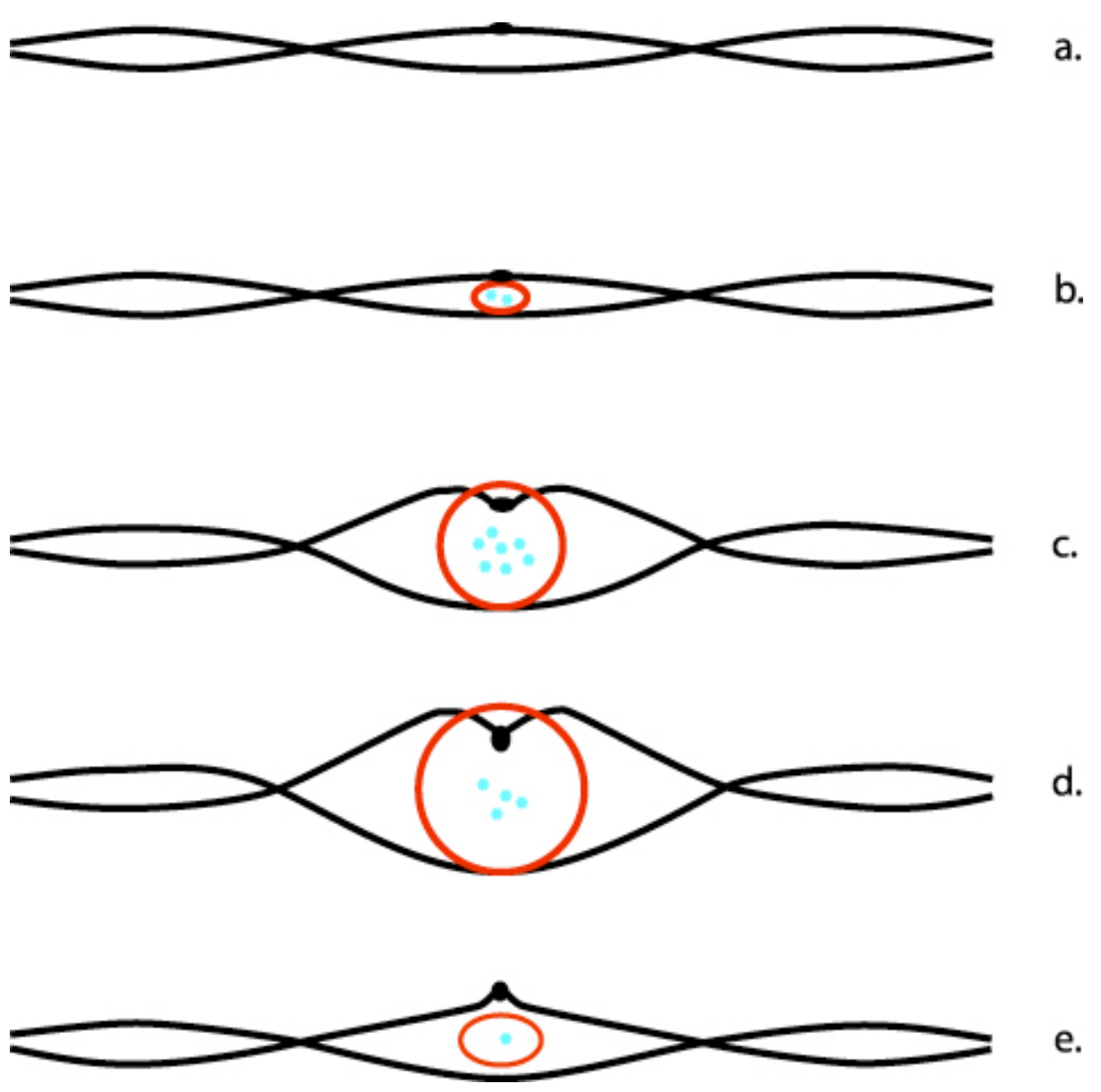

**Figure 14.** Schematic representation of the first stages of the development of an outgrowth according to the model considered. Frame **a)** shows a section of a dusty spiral arm consisting of two dark filaments. **b)** An association of OB stars is just about to be formed in the arm, giving rise to an expanding H II region surrounded by a shell (red oval). **c)** Parts of the dusty arm filaments are swept up by the expanding shell. Only a mass condensation in one of the filaments is lagging behind, causing a V-shaped bend of the filament. If the filament is sufficiently twisted, part of it, close to the point of the V, is wound up into a double helix. In case no double helix is formed during the expansion phase of the H II region, there is a second chance during the recovery phase. **d)** The H II region has reached its maximum size and the OB stars have started to become successively extinct. **e)** Because of its enhanced inertia, the mass condensation requires a longer time to return than the rest of the filament, which may result in a new V-shape of the filament.

The HST images of NGC 4921 and 7049 clearly show that the outgrowths emerge from dark arm filaments. The development can be divided into three different phases; a growth phase, a continued existing under fairly steady conditions, and a regression phase. The growth phase is of course essential for all the outgrowths but it is especially marked for the more extended outgrowths. This phase will be discussed in some more detail in Section 5.4. The second phase represents a more or less long-lasting stationary state before the outgrowth starts to decline.

### *5.3 Folding of young outgrowths*

The study of NGC 4921 and 7049 in Sections 2 and 3 shows that most of the shorter outgrowths appear straight and point roughly radially outwards or inwards from their respective mother filament. However, according to the model proposed the outgrowths can grow out in any direction perpendicular to the mother filament (Section 5.2), thus also out of the plane of the galaxy. The solution to this apparent inconsistency is that an initially oblique outgrowth will not remain so indefinitely since various external forces act upon it. One such force is the gravitational force, which is caused by the dusty arm filament out of which the double helix emanates. This force is in the main oriented radially inwards, towards the filament, perpendicular to it. Another force, which chiefly has its roots in the mass density of the galactic disk, is a gravitational force component, perpendicular to the plane of the galaxy. Furthermore, there is a friction, or damping, force, which increases in strength as the folding double helix approaches the denser gas near the galactic plane.

It should be noted that irrespective of its magnitude, the gravitational force caused by the arm filament cannot appreciably influence the folding-down process of the double helix. The reason for that is that the gravitational force exerted by the filament acts rather much along the double helix and can therefore not give rise any appreciable torque on the double helix.

A force that certainly must affect the folding-down process of the double helix is the gravitational force component, perpendicular to the plane of the galaxy, which is caused by the galaxy. The magnitude of this force is not known in any detail for NGC 4921 and 7049, but just as in Section 5.1, we may gain some guidance from the conditions in the Milky Way Galaxy. In our own galaxy, the perpendicular force component gives rise to oscillatory motions of stars up and down through the galactic plane. Bahcall & Bahcall (1985) estimated the period of the vertical oscillation of the Sun to be in the range 52 – 74 Myr for maximum heights above the galactic plane of 49 – 93 pc. Provided the conditions in NGC 4921 and 7049 are not too different from those in the Milky Way Galaxy, the perpendicular component of the gravitational force should result in folding-down times $\tau_1$, which are of the order of the period of the vertical oscillation or shorter. This means that the double helix should be folded down in times that are shorter than the typical period of galaxy rotation. Depending on the direction in which the outgrowth initially grew out, the outgrowth folded-down will point either outwards away from the centre of the galaxy, or inwards towards the centre.

### *5.4 Further development of the outgrowths*

The motion of stars and interstellar clouds in spiral galaxies cannot be compared with the motion of the planets around the Sun where the orbital velocities decrease outwards as given by Kepler´s third law. Instead, the motion in spiral galaxies is in general fairly well described by a flat rotation curve outside a radius of one to a few kiloparsec. This means that the rotational velocity in most spiral galaxies is mainly independent of the radial distance from the centre and that the rotation period increases linearly with the radius. Velocity measurements have shown that such a rotation is at hand in NGC 7049 (Corsini et al. 2003). As regards NGC 4921, it is harder to measure the rotational velocities and to find the rotation curve since this galaxy is seen nearly face-on. It is, however, reasonable to believe that the rotational velocities are well described by a flat curve also here, excluding the innermost region.

The scrutiny of the dark features in NGC 4921 and 7049 in Sections 2 and 3 reveals that the outgrowths are of very different lengths, ranging from hardly discernible bulges of the dark arm filaments to long lanes comparable in extent with the dusty spiral arms. The short outgrowths being in majority are in general fairly perpendicular to the dusty arm filaments while the longer ones, oriented outwards, tend to bend more and more in the same way as the spiral arms do, thus being trailing. Longer outgrowths oriented inwards also bend but are preceding.

The bending of the outgrowths observed can be readily interpreted within the framework of the present model. According to the model, a double helix starts to develop from the point of a local V-shaped structure in a dusty, arm filament if the filament is sufficiently twisted. In this phase, the double helix protrudes nearly perpendicularly to the dusty arm filament. If new turns are added to the arm filament, the double helix grows out further. As gas is successively brought out into a growing, radially oriented double helix, angular momentum has to be supplied to the structure if it is directed outwards or taken away from it if it is directed inwards. This is equivalent to saying that an azimuthal force must act on the double helix, tending to bend it. If the extension continues, the double helix should finally be bent so much that its outer parts are mainly azimuthally oriented. At this stage, only the gravitational and centrifugal forces, governing the motion of ambient disk stars, act on the outer parts of the double helix. This means, that the outer parts of the outgrowth will more and more adapt the azimutual component of its motion to the flat rotation curve of the galaxy. The implication of that is that the outer parts of the double helix must be further stretched along the azimuthal direction. The deviation is such that an outwardly oriented double helix is trailing while an inwardly oriented double helix is preceding. These kinds of deviation are precisely those observed in NGC 4921 and 7049.

The images of NGC 4921 and 7049 reveal that the outgrowths can change over time. Apart from the fact that stars are found to be formed out of the molecular gas of the outgrowths and must consequently carry away matter from the outgrowths, there are also signs showing that some outgrowths have been larger in previous epochs. In Figure 4, for instance, there is a very dark, V-shaped outgrowth ~ 4″ from the lower edge of the image, with a cloud of rather faint, blue stars stretching more than an arc second outside the point of the V. In the same place a narrow (< 0.″1), dark filament is winding outwards from the point of the V through the star cloud. A reasonable interpretation of this is that the filament represents the remainder of a former double helix filament, attached to the V-shaped filament, out of which the blue stars have been formed. Since the luminous, blue stars visible in the cloud can hardly be older than some 50 Myr, the double helix filament must have declined or eroded in a time that is smaller than that value. Another example of a similar outgrowth with a rudimentary filament and exterior blue stars is found in Figure 3.

To get an idea of the magnitude of the growth times of the longer, bent outgrowths in NGC 4921 and 7049, we shall consider a simple model of the development of such outgrowths. The model is based on two assumptions, which find some support in the discussion above. The first assumption prescribes that the radial velocity component of the head of the outgrowth, $v_{r0}$, is constant during the development of the outgrowth. This is of course a rough approximation but may be considered a first order estimate where $v_{r0}$ stands for a mean value over time of the actual radial growth rate. The second assumption implies that the azimuthal component of the velocity of the head, $v_{\varphi 0}$, is given by the condition that the head of the outgrowth adapts its motion to the flat rotation curve of the galaxy, which means that $v_{\varphi 0}$ is constant, independent of the radial coordinate of the head. The outgrowth is thus growing both radially and azimuthally. During its development, the outgrowth extends outwards from the basic mother filament at radius $r_f$ to the presently observed location of its head at radius $r_h$. The age of the outgrowth can then be expressed as

$$\tau_2 = \frac{r_h - r_f}{v_{r0}} \quad . \tag{4}$$

During that time period, the foot points of the outgrowth move an angle

$$\varphi_f = \int_0^{\tau_2} \omega_f dt = \omega_f \tau_2 \tag{5}$$

as measured from the centre of the galaxy where $\omega_f = v_{\varphi 0}/r_f$ is the angular velocity of the foot points. During the same time period the head moves a different angle

$$\varphi_h = \int_0^{\tau_2} \omega_f \frac{r_f}{r} dt \quad . \qquad (6)$$

Putting $dt = dr/v_{r0}$, we get the azimuthal shift of the head relative to the foot points

$$\varphi_0 = \varphi_f - \varphi_h = \frac{\omega_f \left(r_h - r_f\right)}{v_{r0}} - \int_{r_f}^{r_h} \frac{\omega_f r_f}{v_{r0}} \frac{dr}{r} \quad , \qquad (7)$$

leading to

$$v_{r0} = \frac{v_{\varphi 0}}{\varphi_0} \left( \frac{r_h}{r_f} - 1 - \ln \frac{r_h}{r_f} \right) \quad . \qquad (8)$$

Provided the values of $v_{\varphi 0}$, $r_h$, $r_f$, and $\varphi_0$ can be found, it is possible calculate the magnitudes of $v_{r0}$ and $\tau_2$. Since the outgrowths are split into two or more legs, they generally have more than one foot point. When estimating $\varphi_0$ we use a mean position of the foot points. In order to find the values of $r_h$, $r_f$, and $\varphi_0$ for outgrowths in NGC 7049 we have constructed an artificial, face-on projection of the dust ring in that galaxy (Figure 15) permitting simple measuring. The projection was obtained by assuming the dust ring to be tilted 35° with respect of the line of sight yielding a circular ring. As regards NGC 4921, it naturally presents a face-on view where the values of $r_h$, $r_f$, and $\varphi_0$ can be straightforwardly determined. As indicated above, no rotational velocities are available for NGC 4921. It is, however, reasonable to assume a flat rotation curve and we choose the velocity level $v_{\varphi 0}$ = 250 km s$^{-1}$. In the more inclined NGC 7049, Corsini et al. (2003) have measured radial velocities along the major axis of the galaxy. The absolute values of these levels differ somewhat from one side of the galaxy to the other. After having corrected for the tilt of the ring structure, we adopt the mean value $v_{\varphi 0}$ = 390 km s$^{-1}$. Using the parameters obtained, the values of the quantities $v_{r0}$ and $\tau_2$ have been calculated by means of equations (8) and (4) for some of the longer outgrowths in the two galaxies. The results are presented in Table 1.

Although schematic, the model considered provides a fairly consistent picture of the development of the outgrowths. The radial growth rates are found to be astonishingly similar

**Table 1**

Radial velocity component $v_{r0}$ and age $\tau_2$ derived from the parameters $v_{\varphi 0}$, $r_h$, $r_f$, and $\varphi_0$ for some of the longer outgrowths in NGC 4921 and 7049.

| Outgrowth | Galaxy | $v_{\varphi 0}$ (km s$^{-1}$) | $r_h$ (kpc) | $r_f$ (kpc) | $\varphi_0$ (rad) | $v_{r0}$ (km s$^{-1}$) | $\tau_2$ (Myr) |
|---|---|---|---|---|---|---|---|
| #1 | N 4921 | 250 | 17.3 | 10.7 | 0.94 | 36 | 180 |
| #2 | N 4921 | 250 | 14.4 | 10.7 | 0.40 | 31 | 120 |
| #3 | N 4921 | 250 | 15.2 | 11.6 | 0.33 | 30 | 120 |
| #1 | N 7049 | 390 | 3.7 | 3.2 | 0.32 | 13 | 38 |
| #2 | N 7049 | 390 | 4.3 | 3.4 | 0.31 | 38 | 23 |
| #3 | N 7049 | 390 | 4.5 | 3.4 | 0.47 | 36 | 30 |
| #4 | N 7049 | 390 | 2.2 | 2.8 | - 0.51 | - 21 | 28 |

for all the outgrowths studied. From Table 1 we also observe that the ages within each of the two galaxy samples are rather much the same. In contrast, the ages of the outgrowths in NGC 4921 are decidedly larger than the corresponding ages obtained for NGC 7049. Also the inwardly oriented outgrowth #4 in NGC 7049 comprises values of the parameters $v_{r0}$ and $\tau_2$ that harmonize well with corresponding parameter values of the rest of the NGC 7049 outgrowths considered in Table 1. A further circumstance to note in this connection is that *the ages of the outgrowths derived, are all of the order of, or smaller than, the corresponding rotation periods of the outgrowths.*

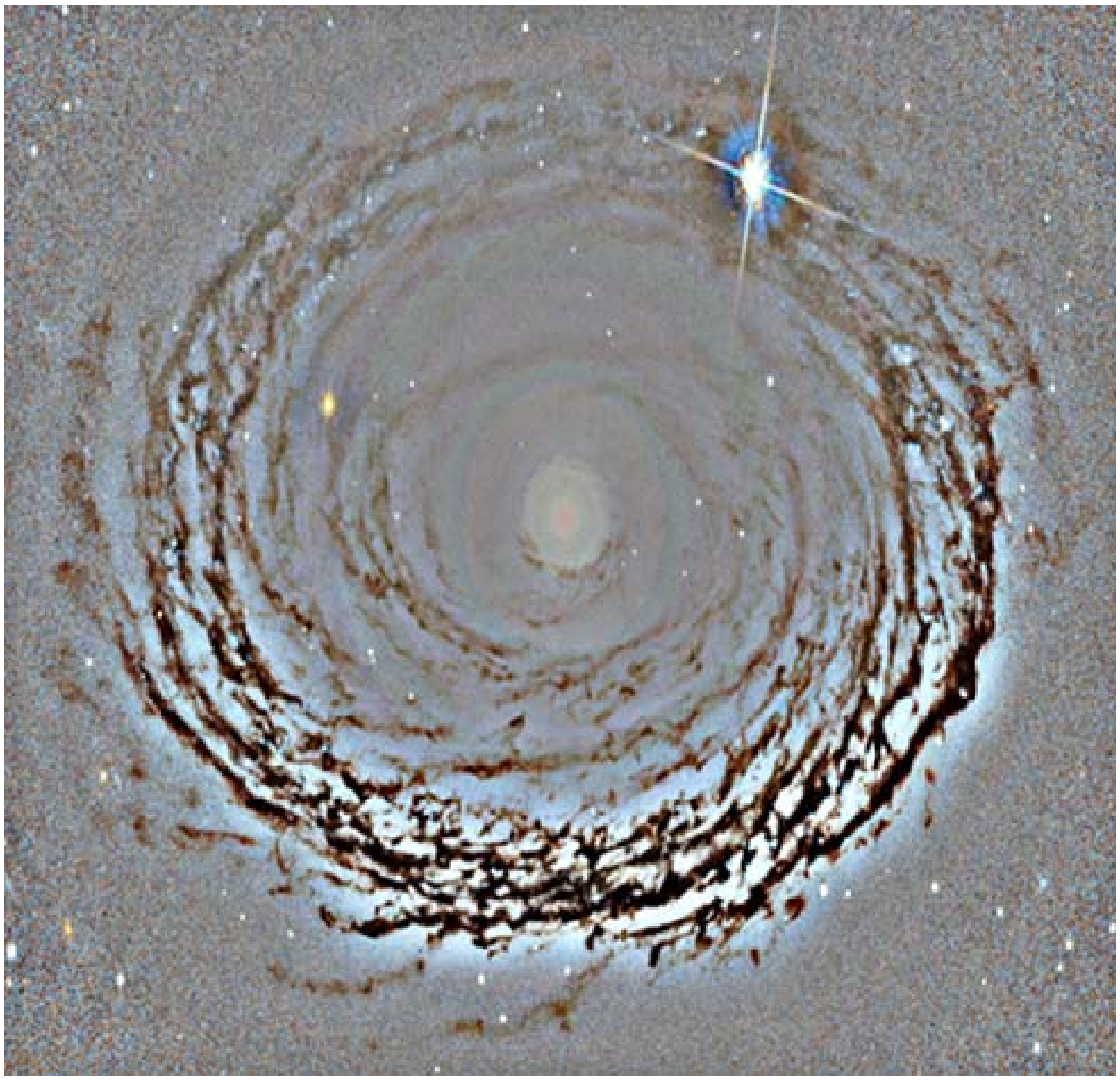

**Figure 15.** An artificially deprojected image, based on the HST image of NGC 7049 in Figure 8, intended to show what the galaxy would have looked like if viewed face-on. It has been assumed that the plane through the dusty ring forms an angle of 35° with the line of sight leading to a circular, deprojected ring. A number of long outgrowths or dust arms can readily be seen to put out from the outer edge of the ring while dragging. Also from the inner rim of the ring similar structures are stretching inwards, but now preceding. If the dark filaments are reasonably cylindrical, as believed, the special projection used should cause the thicknesses of the filaments to appear exaggerated in the upper and lower parts of the ring structure. Such an effect can also be noticed. Likewise, the projection procedure causes an in reality fairly spherical, nuclear region to be vertically drawn out into an ellipse on the image. Just outside the nuclear region several, moderately dark structures exist, which do not look circular. One interpretation of this might be that the structures are in fact truly circular but situated in planes that are tilted with respect to the plane of the dust ring. Another, more likely interpretation, indicated by the shapes of the dark structures, is that they represent the remains of old outgrowths, which have spiralled inwards from the dust ring. In order to bring out the dusty arms and filaments as clearly as possible, also behind the bulge, the image has been treated by a high pass filter.

The model discussed above can also be used to describe the trajectory of the head of an outgrowth with respect to its foot points. We can illuminate this by considering e.g. the prominent outgrowth #1 in NGC 4921. Using the values of $v_{\varphi 0}$, $r_f$, and $v_{r0}$ given in Table 1, for the outgrowth we can by means of equations (4) and (8) calculate the position of the head, $r_h$ and $\varphi_0$, with respect to the footpoints at different times. The positions of the head at the times 0, 30, 60, 90, 120, 150, and 180 Myr after the start of the outgrowth are shown in Figure 16 as small, red crosses. If in addition, all parts of the outgrowth are taken to move according to the same rules as those assumed for the head, the red crosses also mark out the shape of the outgrowth modelled as seen today. As is clear from the figure, there is a fairly good agreement with the appearance of the real outgrowth.

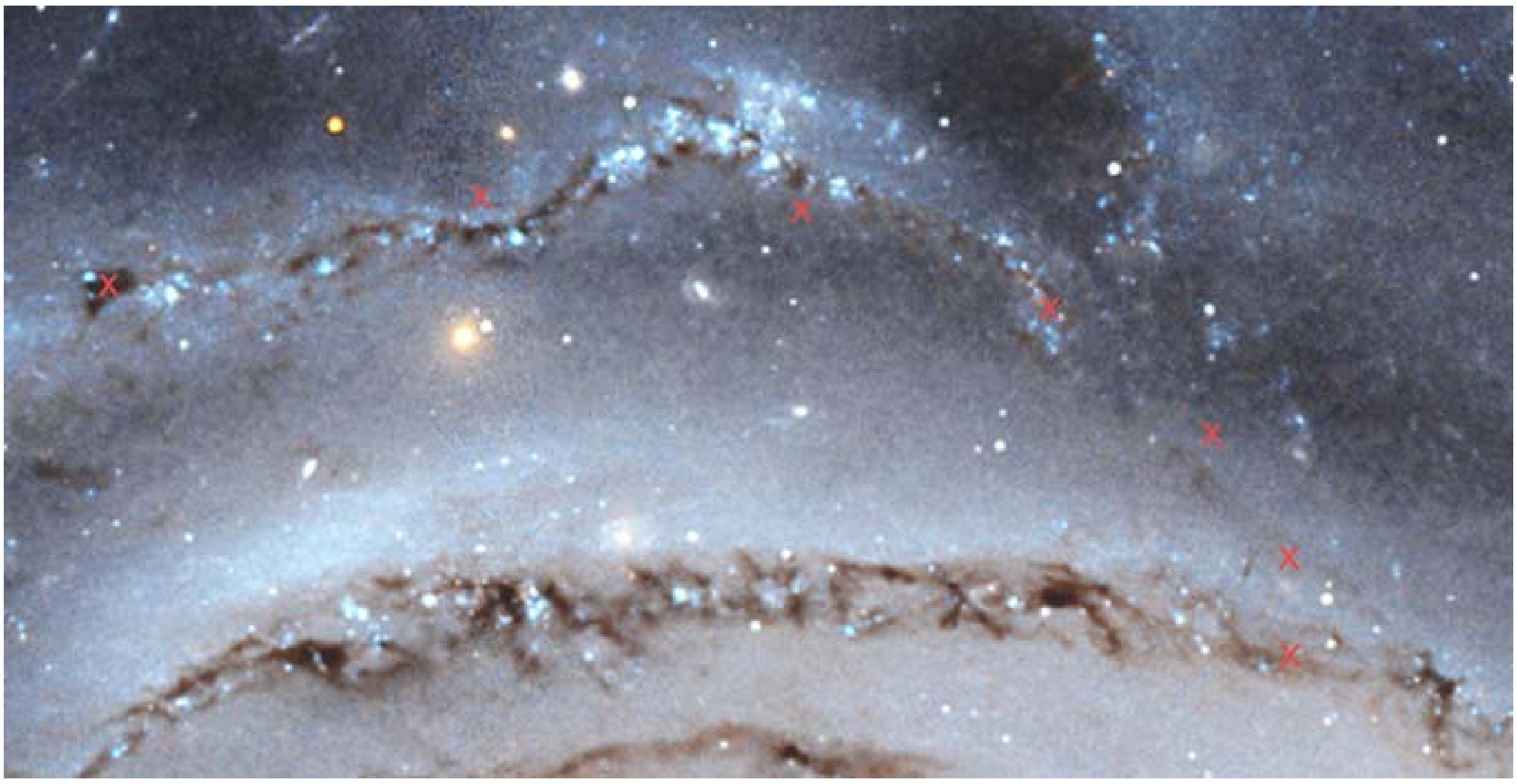

**Figure 16.** Path of the head of the prominent outgrowth #1 in NGC 4921 according to the model discussed in Section 5.4. The path is marked out with small, red crosses illustrating the positions of the head relative to the foot points at 0, 30, 60, 90, 120, 150 and 180 Myr after the outgrowth started to grow. If it is assumed that all parts of the outgrowth move in the same way as the head, the crosses also trace out the present view of the outgrowth as described by the model. Image size: 35″ x 18″.

*5.5 Outwardly and inwardly oriented outgrowths*

A notable property of the outgrowths in as well NGC 4921 as NGC 7049 is that they are oriented both outwards and inwards in their respective systems. Such a geometrical arrangement is neither found with the mammoth trunks in NGC 1316 nor with elephant trunks in H II regions. It is of course of great interest to find out, why the outgrowths behave differently on this point as compared with the mammoth and elephant trunks. The model of the outgrowths here offers a solution. Contrary to the models of the mammoth trunks and elephant trunks, the model of the outgrowths suggests that the outgrowths should be oriented both outwards and inwards. As mentioned in Section 5.2, the V-shaped arm filament, formed by an expanding H II region in a dusty spiral arm, can in principle point in any direction perpendicular to the dusty arm. With this interpretation and the folding-down mechanism discussed above, it is natural to find both outwardly and inwardly oriented outgrowths in the galaxies studied.

## 6 Discussion

### *6.1 Young, blue stars associated with the outgrowths*

A great number of young, blue stars and star clusters are present in both NGC 4921 and 7049. In NGC 7049 the blue stars are mainly scattered among the dusty features in the ring-shaped structure. Since these features are so densely packed, it is hard to link up a given blue star with an outgrowth with certainty. On this point, the situation is quite different in NGC 4921.

In NGC 4921 the arms and outgrowths are more sparsely distributed implying a safer identification of linked objects. Moreover, the blue stars are here mainly situated along the dusty, spiral arms and just outside the heads of outwardly oriented outgrowths (Figures 2 to 5). In some of the outgrowths a small counter-clockwise displacement may be discerned. When present, the displacements are mostly in the same direction as the spiral arms are trailing. By contrast, the inwardly oriented outgrowths are not to the same extent associated with young, blue stars (Figures 7, 8). An exception in this respect is the inwardly oriented outgrowth, shown in Figure 6, where a collection of blue stars is to be seen close to its head.

From the model of the outgrowths one can anticipate that blue, luminous stars should mainly be formed in the heads of the outgrowths. When a twisted arm filament is locally turned into a double helix forming an outgrowth, the two intertwined helical filaments must exert a pressure on each other as a result of magnetic tension. This is analogous to how two intertwined rubber strings influence each other (Figure 13). The enhanced magnetic pressure leads to compression of molecular gas in the outgrowth with subsequent disposition for star formation. Most of all, the magnetic field is bent and tangled in the head of the outgrowth, a place where in addition much gas is initially concentrated according to the model. The highest rate of star formation is therefore expected to take place in the heads of the outgrowths.

In steady outgrowths oriented radially outwards in a galaxy, the rotational velocity about the centre of the galaxy increases linearly outwards with the distance from the centre. As soon as a star is formed in the head of such an outgrowth, it will decouple from the gas. The initial velocity of the star is here somewhat larger than the orbital velocity of ambient disk stars obeying the flat rotation curve. The newborn star can therefore not move in a circular orbit but must after decoupling from the gas, start to move outwards in a slightly elongated orbit. In a coordinate system that co-rotates with the outgrowth and has its origin at the head, the young star will be seen to move slowly outwards from the head while gradually trailing in the same sense as the spiral arms. Calculations show that stars, formed in the head of e.g. a steady, outwardly oriented outgrowth of length 2″ with its base situated at a distance of 25″ from the centre of NGC 4921 (cf. Figure 3), will after 20 Myr and 30 Myr have moved 0.″34 and 0.″73 in the radial direction relative to the head while lagging azimuthally 0.″10 and 0.″32, respectively. Of course, we cannot expect that all star-forming outgrowths are steady. For those, which are not, the motion of the head relative to the base of the outgrowth must influence the positions of young stars at later times.

The blue objects seen outside the outgrowths are thought to consist mainly of O stars and luminous B stars, and clusters and associations of such stars. The upper limits of the lifetimes of such stars are about 30 – 40 Myr. As regards the most massive O stars, their lifetimes are expected to be smaller still, probably well below10 Myr. In this connection, it is of some interest to note that the most intense, blue objects are found next to the heads of the outgrowths while fainter objects exist further out. The detailed interpretation of this is certainly more complicated since it depends on whether the blue objects outside the outgrowths mainly consist of individual, very luminous stars or clusters and associations of somewhat less luminous stars. For comparison, it should be mentioned that an ordinary, main sequence B star can hardly be visible as an individual object on the HST, *ACS* image of NGC 4921.

The blue stars associated with the small outgrowths in Figure 4 show a similar tendency as the blue stars associated with the outgrowths in Figure 3. The more intense blue objects are situated just outside the tips of the outgrowths while fainter objects are found further out. An analogous behaviour may also be at hand for the longer outgrowth in the upper part of the figure. The interpretation is, however, more uncertain here since the blue stars may have been formed in the much larger volume containing the loop-like head. Fainter, diffuse star clouds seem also to be present at distances up to at least 3″ (1.5 kpc) outside the outermost parts of the head.

### *6.2 Rotatory motions within the outgrowths and dusty spiral arms*

According to the model considered, the double helix in a Y-shaped outgrowth extends, at least to begin with, mainly because new turns are added to the double helix. As this occurs, the whole outgrowth has to rotate about its axis just as implied by the filamentary theory of elephant trunks (Section 4.2). In its infancy the Y-shaped outgrowth is thus expected to grow along its own axis while rotating about the same axis. As long as the outgrowth extends, mainly because new turns are added to the double helix, we can estimate its rotational velocity from the helix angle of the double helix and the growth rate along the axis. If the helix angle in the outer parts of the double helix is ~ 45° as observed (Section 3), the rotational velocity of these parts is expected to be of the same magnitude as the axial (or radial) growth rate $v_{r0}$ (cf. Table 1). In later stages of the development the extension of the outgrowths may depend also on other mechanisms than the addition of new turns to the double helix. Under such circumstances the rotational velocity could very well be slower than the axial motion. It is even possible that the extension of the outgrowth ceases or is turned negative, indicating that the rotation has stopped or reversed.

In view of these expectations, it is of course of obvious interest to find out if there is any observational support for rotatory motions in the outgrowths. The detailed HST images shown above can here offer some aid. As already hinted at, the twined geometries observed in many of the outgrowths in NGC 4921 and 7049, together with the frozen-in conditions of the magnetic field, constitute strong evidence of that rotatory motions have actually taken place in the outgrowths. Interestingly, also the dusty spiral arms in the two galaxies are made up of dark filaments, which in long pieces appear to be twined about each other (e.g. Figures 2, 10) indicating that rotatory motions have been in action in the arms as well.

Another phenomenon that may have some bearing on a rotation of dusty arm filaments is the so-called *rolling motion phenomenon* in the spiral arms of our own galaxy. Oort (1962a) first noticed the phenomenon finding that the motions of gas in the spiral arms of the Milky Way Galaxy are not symmetrical with respect to the galactic plane. Instead, the radial velocities are systematically larger on one side of the plane than on the other side for the same galactic longitude. On the basis of H I data Feitzinger & Spicker (1985) could eliminate an interpretation of the rolling motion phenomenon based on differential rotation and the bending of the galactic plane (Yuan & Wallace 1973). Instead, they were able to demonstrate the existence of real, rolling motions in the spiral arms with typical velocity gradients of ± 20 km $s^{-1}$ $kpc^{-1}$ and absolute, maximum values a few times larger. In addition to this, the phenomenon has also been observed in the CO lines (Wouterloot 1981) showing that rolling motions exist also in the dusty arms of the Milky Way Galaxy.

### *6.3 Outgrowths and dust arms in NGC 4921 and 7049*

As demonstrated in Section 2, the outgrowths in NGC 4921 are of different sizes ranging from small, barely discernible bulges on the dusty arm filaments to the prominent outgrowth #1 shown in the upper half of Figure 2. With a length of more than 17 kpc, the latter one is in several respects strikingly similar to the two dark, spiral arms in the lower half of the same figure. Only the fact that the prominent outgrowth is connected through two or more legs to one of the arms in the figure suggests that we are here dealing with an outgrowth. Taking the fact that the legs in this outgrowth are fairly dilute into account, we conclude that it is not a simple task to distinguish a well-developed outgrowth from a genuine, dusty, spiral arm.

In NGC 7049 there are a number of dark and comparatively long features, which protrude outwards from the ring structure while lagging behind (e.g. #1, #2, and #3 in Figures 9, 15). Whether these features represent outgrowths or genuine, dusty, spiral arms is not easy to tell. Without knowing anything about the outgrowths, one would probably classify them as spiral arms. However, if we look somewhat closer to these features, we find that most of them show all the marks of the outgrowths. They are dark, they are mostly double, many of them appear to be twisted, they are often closed by a loop at their tips, and they connect by means of legs to already existing filaments in the ring although it may in some cases be hard to follow the details.

The find that the prominent outgrowth in NGC 4921 and several of the longer, lagging outgrowths outside the ring structure in NGC 7049 can be mixed up with dusty spiral arms naturally addresses an important question. Could it be so, that there is only a vague, physical distinction between well-developed outgrowths and genuine dust arms? The existence of the prominent outgrowth in NGC 4921 and the long, lagging outgrowths outside the ring in NGC 7049 indicates that this could very well be the case. New, dusty arms may be formed as the result of the growth of initially small outgrowths. The model suggests that, as gas is brought out to a growing outgrowth in the form of a double helix, the content of gas in the mother arm should decrease (Section 5.5). In other words, the growth of a new arm should lead to the decline and fading of the old arm.

The evolution from an outgrowth to a dusty, spiral arm should not, in principle, differ from the evolution of an ordinary outgrowth – it is only more powerful. Gas is drawn out from the dusty mother-arm filament to the new arm through a more pronounced and perhaps prolonged growth of the double helix where at later stages azimuthal stretching substantially contributes to the extension of the new arm.

In addition to the long and obvious candidates for the dusty arms mentioned above, there also exist outgrowths of intermediate lengths, which may later develop into genuine arms (e.g. the outgrowths #2, #3 in Figure 5). All outgrowths are, however, not expected to develop into spiral arms. On the contrary, one should anticipate that only a small fraction of the outgrowths is subject to such an extreme growth. Firstly, the winding up of a filament into a double helix may after some time be turned into its reverse, leading to stagnation or even decline of the outgrowth. Secondly, the limited gas content in the old, mother arm can probably sustain only a very small number of new arms – perhaps just one. Hence, only the most vigorous outgrowths can be expected to develop into full-fledged arms. In the following, we shall discuss some implications of such a development and investigate how far the model can serve.

### *6.4 Outgrowths in other galaxies?*

Above, we have in some detail studied the outgrowths and dusty arms in the two spiral galaxies, NGC 4921 and 7049. A question of obvious interest is of course if these outgrowths are unique or whether similar features are present also in other spiral galaxies. As regards the dusty arms, it has for a long time been known that such arms exist in almost all well-developed, spiral galaxies. Generally, they are found near the inner rims of bright, spiral arms (Sandage 1961; Lynds 1970; Weaver 1970). Mostly, the dust arms consist of two or more, relatively thin filaments, which in places seem to be twined around each other and, in other places again, run side by side. Numerous examples of such configurations may be found e.g. in The Hubble Atlas of Galaxies (Sandage 1961) and in HST images of spiral galaxies as well. However, the structures of dust arms in most spiral galaxies are generally more disturbed and complex than they are in NGC 4921. In this respect, the latter galaxy therefore constitutes a favourable object, which is simpler to study and interpret than most other spirals.

Just as the outgrowths are protruding from the dust arms in NGC 4921 and 7049, there are also dark structures extending from the dust arms in many other spiral galaxies (Lynds 1970; Weaver 1970). For a long time, such structures have been termed *feathers* on the proposal of Lynds. Using photographic plates from the Mount Wilson and Palomar observatories, both Lynds and Weaver noted the presence of thin extinction features stretching outwards with large pitch angles from the dusty arms in spiral galaxies. In a thorough study of several spirals, La Vigne, Vogel & Ostriker (2006) later pointed out that feathers are most common in Sb–Sc galaxies. Particularly well-developed feathers are found in grand design spirals like NGC 5194 (M 51), NGC 5236 (M 83), and NGC 5457 (M 101). Studies of images of spiral arms captured by the HST reveal that feathers are made up of dark filaments, which in places show clear signs of being twisted about each other. More often, however, the structures are fairly disordered and, hence, harder to interpret. Usually, feathers are connected to dusty arm filaments by means of two or more filamentary legs, just as the outgrowths are. If sufficiently extended, feathers bend in the same way as spiral arms do. Detailed HST images show that also inwardly oriented dust features are present in spiral galaxies, although to a less extent than the outwardly oriented ones. Just as there are long and prominent outgrowths in NGC 4921 and 7049, there also exist very long feathers in other spiral galaxies, some of which can be compared to dusty arms.

Another property, common to feathers and outgrowths, is their close association with hot, young stars. It was early noted by Weaver (1970) that bright chains of OB stars and H II regions, so called *spurs*, extend outwards from the arms, into the interarm region (here we stick to the definitions of the terms spur and feather given by La Vigne, Vogel & Ostriker 2006). Piddington (1973) drew attention to the fact that spurs are often associated with feathers and Elmegreen (1980) further stressed this connection. In galaxies like NGC 5194 and NGC 5457, the presence of hot, blue stars in, and just outside, feathers is particularly evident. Generally, the young stars form more or less extensive clouds outside the dust arms. Together with dust arms, feathers constitute important sources of young, bright stars in galaxies (Balbus 1988; Kim & Ostriker 2002; La Vigne, Vogel & Ostriker 2006). In fact, the young stars are essential for our immediate impression of normal, well-developed spiral galaxies since they contribute substantially to the bright arms. Similarly, one may interpret the blue stars found outside the dust arms in NGC 4921 as rudiments of bright arms. In conformity with the situation in NGC 4921, young stars in spurs are often somewhat displaced, both radially and azimuthally, so that they tend to be trailing outside the feathers.

From the above discussion, it is clear that there are considerable similarities between the dark outgrowths in NGC 4921 and 7049 on the one hand, and feathers in other spiral galaxies on the other hand. There are thus good reasons to believe that the two kinds of object are closely related to one another. Hence, important parts of the model of the outgrowths may also be applicable to feathers. If so, feathers too may contribute to the formation of new dust arms in other spiral galaxies.

In recent decades, a number of theoretical studies of feathers have been performed. In a seminal work, based on time-dependent, numerical, MHD simulations of the interaction of a magnetized interstellar medium with a differentially rotating stellar spiral arm, Kim and Ostriker (2002) found that gaseous structures, roughly resembling feathers, can emerge from self-gravitating perturbations. They interpreted the structures developed in terms of the magneto-Jeans instability. More recently, several papers on the formation of feathers, mainly based on MHD, self-gravity simulations, have been published with ever increasing numerical sophistication (e.g. Chakrabarti, Laughlin & Shu 2003; Kim & Ostriker 2006; Shetty & Ostriker 2006; Dobbs & Bonnell 2006; Dobbs 2011; Lee & Shu 2012). Many of the models proposed are capable of giving a fairly good description of the general structure of feathers and their mutual locations but are less detailed in accounting for the characteristics of their internal structure. In this respect, the outgrowth model of feathers may have a point.

From the previous sections it is clear that the outgrowth model of feathers can naturally give an account of the filamentary character of feathers with their often very clear attachments to the dusty arm filaments by means of two or more filamentary legs. To the preferences of the outgrowth model one may also add that the model can naturally explain the twisted character seen in some feathers. The twisting of feathers is certainly not so marked as with many of the outgrowths studied in NGC 4921 and 7049. The reason for that may be sought in the fact that the outgrowths represent very simple and comparatively undisturbed structures while most feathers have been subject to considerable shuffling and mixing. A further distinctive property of the model is that it can give an account of why dusty structures can point inwards in the galaxy while preceding. Even if the outgrowth model can thus offer answers to some of the most basic questions regarding the feathers, one should not forget that there still remain many details to elucidate before a reasonably complete theory is achieved.

### *6.5 The winding problem of spiral arms*

For decades, the winding problem of spiral arms has both hampered and inspired the discussion on how spiral arms are formed and develop. Since most spiral galaxies rotate differentially with a decreasing angular velocity outwards, one should expect that their arms would be gradually wound up into ever-tighter spirals (Oort 1962b). However, this is not what is generally observed. A possible way out of this dilemma has been offered by the density-wave theory proposed by Lin and Shu (1964, 1966). The theory, which is to some extent related to earlier works by Lindblad (1951, 1961, 1963) and Lindblad (1960), suggests that spiral arms are constituted by waves of excess density, in which stars and gas clouds are temporarily crowded together. The waves move around the galaxy with velocities that usually differ from the velocities of ambient disk stars and gas clouds. Hence, in this view spiral arms must be regarded as patterns of enhanced density rather than as structures containing a specific collection of stars and gas clouds. When orbiting a spiral galaxy, stars and gas clouds are thought to pass straight through the arms in comparatively short times. This pattern of motion obviously differs very much from the kind of motion considered above for the outgrowths and dusty spiral arms.

The model of the outgrowths offers a new way of approaching the winding problem of spiral arms. As discussed above, the model suggests that new arms can be formed at the expense of old arms in time periods $\tau_2$ that are shorter than the period of rotation of the galaxy. If this is what happens, there may simply not be time enough for the arms to be wound up into tight spirals. Likewise, the folding-down times found are relatively short. The fact that star formation in the longer outgrowths in NGC 4921 and 7049 is still going on, also lends some support to the view that even the longer outgrowths are relatively young objects.

Additional arguments, indicating that the spiral structure may change comparatively rapidly in galaxies, have been given by among others Merrifield, Rand & Meidt (2006).

### *6.6 Properties, predictions, and further observations*

Above, we have discussed two kinds of object that may at the first glance appear to be rather different – the outgrowths and dusty spiral arms. A closer study has, however, indicated that the outgrowths and dusty spiral arms are closely related, which in turn has suggested that basically the same model can be applied to the two kinds of object. As is the case for any model, it is also here urgent to try making testable predictions in order to find further observational support for the model or arguments against it. When discussing this matter it is appropriate to treat the two applications of the model separately.

Considering the outgrowths first, we would like to draw attention to a few important properties of the model, which may be tested. First, i) the model suggests that the two dark, intertwined pieces of the filaments, which make up the stem of the Y-shaped outgrowth, should contain oppositely directed magnetic fields. According to the model, the magnetic fluxes over the cross-sections of the filaments should be of equal magnitude but have opposite signs. If each of the filaments consists of two or more sub-filaments the sum of the fluxes in the sub-filaments should instead be taken into account. A second prediction ii) concerns a possible rotation of twined outgrowths about their long axes. Just as elephant trunks in H II regions have been found to rotate, we expect that most outgrowths should also rotate about their axes, at least for some period of time during their evolution (Section 6.2). For short outgrowths the rotational velocities should be of the same order of magnitude as $v_{r0}$. Regarding the values shown in Table 1 as typical, the rotational velocities should be of the order 20 – 40 km s$^{-1}$. A third item iii) refers to the young, blue stars found outside most of the outgrowths in NGC 4921. As pointed out in Section 6.1, the heads of the outgrowths with their compressed gas and tangled magnetic fields constitute excellent breeding grounds for new stars. As soon as a new star has been formed in the head of a steady outgrowth, it will decouple from the gas in the outgrowth and slowly move outwards while gradually trailing. It would be useful to observe more systematically if and how the colour index of such stars depends on the distances of the stars from the head of the outgrowth. From the model it is expected that the most luminous, and hence bluest, stars should be found only close to the head while somewhat cooler stars should have had time to migrate further out (cf. Section 6.1). There are hints of such a behaviour in Tikhonov & Galazutdinova (2011) where it is indicated in their Figure 9 that the fainter stars outside some of the outgrowths in NGC 4921 are of a greater age than the more luminous stars. However, the stars considered are anonymous belonging to several different outgrowths without any information on their distances outside their respective outgrowth. It would be of great interest to see whether a distance-age relation for stars outside one or more of the outgrowths could be obtained from this or some other investigation.

In order for the model of the outgrowths to be applicable also to the dusty spiral arms, the arms must possess some specific qualities. The following points are essential for the applicability of the model in this respect. First, i) we may conclude from the model that since the dusty spiral arm is made up of a dark, filamentary loop that may be very elongated, the arm must at any place consist of at least two dark filaments. More than two filaments may be present if the basic filament is composed of two or more sub-filaments. Secondly, ii) the filamentary parts making up the elongated loop should, at least here and there, be twisted around each other, thus forming at least a rudiment of a double helix. In places, however, the arm filaments may be so tightly wound around each other that they may be mistaken for a single filament (e.g. see Figure 2). Thirdly, iii) the model suggests that as the young, dusty arm starts to develop, it must rotate about its axis at the same time as its length increases. Radial growth rates of the order 20 – 40 km $s^{-1}$ have been deduced for some of the arm-like outgrowths in NGC 4921 and 7949 (Table 1). The azimuthal growth rate may be even larger – about 80 km $s^{-1}$ for the prominent outgrowth (#1) in NGC 4921. This growth rate is, however, not a velocity on equal footing with the radial growth rate. Rather, it represents a stretching and an adaptation to the flat rotation curve. In the fourth place, iv) the model suggests that the two, basic filaments forming the dusty spiral arm must carry magnetic fluxes that are of equal magnitudes but have opposite signs. As a consequence of that, the total flux across the two filamentary parts should be zero, just as with the outgrowths. This condition may seem fairly natural when applied to the outgrowths and arm-like features in NGC 4921 and 7049 having closing loops at their tips but can perhaps appear less evident for genuine, dusty spiral arms. One should, however, observe that the magnetic fields aimed at, refer to the dusty arms only, not to the surrounding regions where different conditions may prevail.

Although the model of the outgrowths is able to give a good account of a number of characteristics of the individual spiral arm, it is not certain that it alone can properly describe the beautiful symmetry of spiral arms presented by many galaxies. A difficulty in this connection is that the magnetic forces may not be sufficiently far-reaching to co-ordinate the positions of the arms unless the co-ordination occurs close to the centre of the galaxy. If that is not the case, some more global force – preferentially the gravitational force – probably has to assist in the creation and maintenance of symmetry. In this connection one should keep in mind, however, that there also exist many spiral galaxies within which symmetry does not prevail. Still, the arms of such galaxies are often organized into graceful spirals. From that we may conclude that co-ordinating forces are here either inefficient or may not have had sufficient time to act, providing scope for a more direct application of the model of the outgrowths.

## 7 Conclusions

The presence of protruding, dusty mammoth trunks in the odd-looking galaxy NGC 1316 has raised the question whether similar structures may also exist in other, more regular galaxies. To illuminate this issue we have in some detail studied the morphology of the dust features in the two spiral galaxies NGC 4921 and 7049 using HST archival images. Although both galaxies are spirals containing dusty arms, their appearances are quite different. The dust arms in NGC 4921 show a very simple pattern while the corresponding arms in NGC 7049 present a great number of filamentary features, mainly packed together in a ring-shaped structure running around the centre of the galaxy. Generally, the dust arms are made up of two or more dark filaments, which in places appear to be twisted about one another.

The study reveals that both galaxies contain a great number of dark outgrowths emanating from dusty arm filaments, just as NGC 1316. A notable observation is that the outgrowths are oriented both outwards and inwards in the two galaxies. Mostly, the outgrowths are made up of twined filaments, which connect to the dusty arm filaments through two or more filamentary legs. In some cases, the outgrowths are V-shaped with the points oriented away from the arm filaments. The length of the outgrowths varies considerably, from small bulges to long structures protruding more than 10 kpc.

The outgrowths in NGC 4921 and 7049 show considerable similarities to elephant trunks in H II regions in our own galaxy and mammoth trunks in NGC 1316. Markedly filamentary and often twisted, they all contain dusty molecular gas and connect to dark mother filaments by means of at least two legs. The outgrowths also differ from the elephant and mammoth trunks on a few, but important, points. In H II regions and NGC 1316, on the one hand, the two kinds of trunk are oriented inwards, although with substantial scattering. In NGC 4921 and 7049, on the other hand, the outgrowths are oriented both inwards and outwards in the galaxies with the latter category dominating. Another difference is that elephant trunks are generally much smaller than the mammoth trunks and outgrowths. The similarities indicate that there should be a clear relationship between the mechanisms leading to the protruding structures while the differences imply that the mechanisms cannot be identical.

Founded on basic properties of the outgrowths, such as their filamentary character and twisting, a model of the outgrowths is proposed. Having several elements in common with a recent theory of elephant trunks, the model suggests that a magnetized filament can under certain conditions be locally transformed into a V-shaped structure or double helix putting out from the filament. For the double helix to be formed, it is required that the filament with its frozen-in magnetic field lines is twisted beyond a certain critical limit. The magnetized filament is identified with a dusty, spiral-arm filament while the double helix or V-shaped structure represents the outgrowth. The model is thus capable of accounting for both the filamentary nature of the outgrowths and their twisting. According to the model, the development of the double helix takes place in two stages, which may, or may not be separated in time. In the first stage a dark arm filament containing a mass condensation is drawn out to V-shape when an expanding H II region is formed in the dusty arm. In the second stage, if it occurs at all, the magnetic field inside the filament is twisted beyond a critical limit leading to that the filamentary parts next to the point of the V are transformed into a double helix. The result is a Y-shaped outgrowth. If the twisting of the filament is not sufficiently large, the filament forms a V-shaped outgrowth.

The model proposed for the outgrowths has several properties in common with a recent model of the mammoth trunks in NGC 1316. Most important of these properties is that both models rely on the double helix mechanism. However, the two models also differ from one another. The model of the mammoth trunks restricts the trunks to be oriented only inwards while the model of the outgrowths permits them to be oriented both outwards and inwards.

According to the model of the outgrowths, an outgrowth in the making can protrude in any direction perpendicular to the dusty arm filament from which it evolves. This means that also oblique outgrowths, putting out of the plane of a galaxy, can be formed. Such outgrowths are, however, expected to be quickly folded down towards the plane of the galaxy. The folding-down is mainly accomplished by a component of the gravitational force, perpendicular to the plane of the galaxy. The time periods estimated for the folding-down process are distinctly shorter than the typical rotation periods of the galaxies.

In the beginning of its evolution the Y-shaped outgrowth protrudes outwards or inwards in the galaxy mainly through the addition of new turns to the double helix. Later on, if the outgrowth develops further, it is also drawn out in the azimuthal direction by the differential rotation. This agrees well with the observed behaviour of the outgrowths. According to a sub-model describing the evolution of the outgrowths quantitatively, the growth is rapid in the sense that the time periods required for the extension of even the longest outgrowths in NGC 4921 and 7049 are of the same order, or shorter than, the periods of rotation of the outgrowths about the centres of the galaxies. It is noteworthy that the sub-model is also capable of accounting for the general shape of the outgrowths.

When a dusty arm filament is locally wound up into a double helix forming an outgrowth, the molecular gas in it is compressed, promoting star formation. The star-forming process should be most efficient in the head of the outgrowth where the magnetic field is most tangled and where, in addition, a mass condensation usually exists. In a steady, outwardly oriented outgrowth, the outer parts of the outgrowth must move with velocities that are somewhat larger than the orbital velocities of ambient disk stars. A new star, formed in the head of such an outgrowth, should therefore move a bit faster than ambient disk stars. As a consequence of that, the new star will, after decoupling from the gas, slowly move outwards relative to the head of the outgrowth while gradually trailing. In case of a more dynamic outgrowth, the motion of a newly formed star is different.

One of the outgrowths – the longest and most prominent one in NGC 4921 – shows a remarkable similarity to the dusty, spiral arms in the galaxy. The outgrowth is clearly filamentary with at least two narrow, dark filaments, which in places spiral around one another. What distinguishes the prominent outgrowth from a genuine, dusty, spiral arm in the galaxy is that it is connected to one of the arms by means of at least two legs. The outgrowth is mainly azimuthally oriented with a length of more than 17 kpc, roughly corresponding to its radial distance from the centre of the galaxy. Young blue stars are present along the outgrowth and just outside it. Also in NGC 7049 one finds outgrowths that very much resemble spiral arms. In fact, it is even harder in that galaxy to judge whether a dusty feature putting out from the ring-like structure should be classified as an outgrowth or a spiral arm.

The presence of long, arm-like outgrowths in both NGC 4921 and NGC 7049 shows that the distinction between outgrowths and dusty spiral arms is sometimes minute or even missing. The arm-like appearance of some of the outgrowths in the two galaxies therefore raises the question whether outgrowths can be transformed into new, dusty, spiral arms. There seems not to be any fundamental property of the model proposed that prevents an outgrowth from growing so large that it rivals the spiral arms. One has therefore to face the possibility that some outgrowths in NGC 4921 and 7049 can be, or have already been, transformed into genuine spiral arms. Hence, it is of obvious interest to consider what such a transformation could imply. During the growth phase, molecular gas is brought from the mother arm to the new arm. The time period needed for this process is estimated to be of the same order as, or shorter than, the rotation period of the arm in the galaxy. The process means that as a new arm emerges and flourishes, the old mother arm shrinks and fades. The mother arm and new arm are, however, not expected to form a completely closed system constituting perfect, communicating vessels. Newly formed stars escape from the dusty arms thus carrying away gas while fresh gas may be sucked in to the system from outside.

A special kind of dark feature, which is observed in many spiral galaxies and very much resembles the outgrowths in NGC 4921 and 7049, is the feather. Like the outgrowths in NGC 4921 and 7049, feathers are dusty, filamentary and often somewhat twined structures that protrude from dark arm filaments at considerable pitch angles. In general, they are connected to the filaments by means of two or more filamentary legs. Feathers are found to be especially common in Sb–Sc galaxies and grand design spirals. Since feathers are in many respects so similar to the outgrowths, it is legitimate to suppose that they too can be well described by the model of the outgrowths.

Closely coupled with feathers are spurs, which are seen to jut out at large angles from dusty spiral arms. Spurs are made up of bright OB stars and H II regions and constitute a significant part of the bright arms outside the dusty arms in most spiral galaxies. It is of some interest to note, that the agglomerations of bright OB stars outside the outgrowths in NGC 4921 may be interpreted as mini-forms of spurs.

There are many examples of long and well-developed feathers in spiral galaxies, which in important respects resemble dusty spiral arms. It is therefore reasonable to consider the possibility that feathers, like the outgrowths in NGC 4921 and 7049, can occasionally grow out into new spiral arms at the expense of the old arms. The implications of such a scheme are significant. For instance, most dust arms should be made up of two or more dark filaments. In addition to that, the formation of new arms may offer a new approach to the old, winding problem of spiral arms. If new arms are successively formed out of old arms on time-scales that are of the same order as, or shorter than, the rotation period of the arms, a continuing winding-up of the arms into ever-tighter spirals may never take place.

**Acknowledgements** I wish to thank an unknown referee for many valuable suggestions. I am also much indebted to R. Geissinger, Remseck, Germany for permitting me to use his image on IC 1396, M. Raadu for helpful comments, and A. Brink for kind assistance with the figures. My thanks furthermore go to NASA, ESA, and K. Cook (Lawrence Livermore National Laboratory, USA), NASA, ESA, and W. Harris (McMaster University, Ontario, Canada).